\documentclass[11pt]{article}
\usepackage{amsfonts}
\usepackage{amssymb}
\usepackage{amsmath}
\usepackage{geometry}
\usepackage{graphicx}
\usepackage{float}
\usepackage{color}

\usepackage[small,bf]{caption}
\newtheorem{theorem}{Theorem}
\newtheorem{lemma}[theorem]{Lemma}
\newtheorem{remark}[theorem]{Remark}
\newtheorem{proposition}[theorem]{Proposition}
\newtheorem{corollary}[theorem]{Corollary}
\newtheorem{definition}[theorem]{Definition}

\begin{document}

\title{Spectral properties of a 2D scalar wave equation with \\
1D-periodic coefficients: application to SH elastic waves}


\author{ A.A. Kutsenko$^{a}$, A.L. Shuvalov$^{a}$, A.N. Norris$^{b}$,
            O. Poncelet$^{a}$  \\ \\
            $^{a}$ Institut de M\'{e}canique et d'Ing\'{e}nierie de Bordeaux,
            \\ 
             Universit\'{e} de Bordeaux,
                UMR CNRS 5295, Talence 33405, France. 
            \\  $^{b}$   Department of Mechanical and Aerospace Engineering, 
            \\
            Rutgers University, Piscataway, NJ 08854-8058, USA}

\maketitle

\begin{abstract}

The paper provides a rigorous analysis of the dispersion spectrum of SH
(shear horizontal) elastic waves in periodically stratified solids. The
problem consists of an ordinary differential wave equation with periodic
coefficients, which involves two free parameters $\omega $ (the frequency)
and $k$ (the wavenumber in the direction orthogonal to the axis of
periodicity). Solutions of this equation satisfy a quasi-periodic boundary
condition which yields the Floquet parameter $K$. The resulting dispersion
surface $\omega (K,k)$ may be characterized through its cuts at constant values of $K,~k$ and $\omega $ that define the passband (real $K$) and stopband areas,
the Floquet branches and the isofrequency curves, respectively. The paper
combines complementary approaches based on eigenvalue problems and on the
monodromy matrix $\mathbf{M}$. The pivotal object is the Lyapunov function $\Delta \left( \omega ^{2},k^{2}\right) \equiv \frac{1}{2}\mathrm{trace}\mathbf{M}=\cos K$ which is generalized as a function of two variables. Its analytical properties, asymptotics and bounds are examined and an explicit
form of its derivatives obtained. Attention is given to the special case of
a zero-width stopband. These ingredients are used to analyze the cuts of the
surface $\omega (K,k).$ The derivatives of the functions $\omega (k)$ at
fixed $K$ and $\omega (K)$ at fixed $k$ and of the function $K(k)$ at fixed $\omega $ are described in detail. The curves $\omega (k)$ at fixed $K$ are
shown to be monotonic for real $K,$ while they may be looped for complex $K$
(i.e. in the stopband areas). The convexity of the closed (first) real
isofrequency curve $K(k)$ is proved thus ruling out low-frequency caustics
of group velocity. The results are relevant to the broad area of
applicability of ordinary differential equation for scalar waves in 1D
phononic (solid or fluid) and photonic crystals.

\end{abstract}


\section{Introduction}

The wave equation with periodic coefficients is ubiquitous in physics and engineering. Its applications in acoustics of solids have gained a new momentum since the introduction of artificial periodic materials such as phononic crystals. A common mathematical framework is the Floquet-Bloch theory of partial differential equations with periodic coefficients \cite{Kuchment}. It does not however yield many explicit results for the general case of 2D or 3D periodicity and vector waves. The notable exception allowing an explicit analysis is the case of 1D periodicity and scalar waves which is governed by Hill's equation \cite{MW}. The spectral properties of Hill's equation are very well understood for the situation where the wave propagates along some fixed direction (parallel to the periodicity axis or not). This case implies a single spectral parameter. The objective of the present paper is to take on a broader perspective of arbitrary (2D) propagation of scalar waves in 1D periodic media. This setup implicates dependence on two spectral parameters and thus leads to more elaborate wave spectral properties.  The  specific problem to be addressed is described next.

Consider SH  (shear horizontal) wave motion of the form $u_{z}\left( x,y,t\right)
=U(y)\exp \left[ i\left( kx-\omega t\right) \right] $ which travels in the
symmetry plane $XY$ of a stratified monoclinic elastic solid with periodic
density $\rho \left( y\right) =\rho \left( y+T\right) $ and stiffness $%
c_{ijkl}\left( y\right) =c_{ijkl}\left( y+T\right) $. The elastodynamic 
equation yields a second-order ordinary differential equation for the
amplitude $U(y),$
\begin{equation}
\partial _{j}\left( c_{ijkl}\partial _{l}u_{k}\right) =\rho \ddot{u}_{i}\
\Rightarrow \left( c_{44}U^{\prime }+ikc_{45}U\right) ^{\prime }+ik\left(
c_{45}U^{\prime }+ikc_{55}U\right) =-\rho \omega ^{2}U,  \label{1.0}
\end{equation}
where $\partial _{1}\equiv \partial /\partial x,\ \partial _{2}\equiv
\partial /\partial y,\ ^{\prime }\equiv d/dy$ and 
Voigt's indices $4=yz,$ $5=xz$ are used
\cite{AuldI}. It is convenient to pass from $U$ to $u=Ue^{i\varphi }$ with $%
\varphi \left( y\right) =ik\int^{y}\left( c_{45}/c_{44}\right) \mathrm{d}y$
which  reduces (\ref{1.0})$_{2}$ to the Sturm-Liouville form
\begin{equation}
\left( \mu _{1}(y)u^{\prime }(y)\right) ^{\prime }-k^{2}\mu
_{2}(y)u(y)=-\omega ^{2}\rho (y)u(y),  \label{1}
\end{equation}%
where $\mu _{1}=c_{44}$ and $\mu _{2}=c_{55}-c_{45}^{2}/c_{44}$ denote the  shear moduli. 
Equation (\ref{1}) is the object of our study.  The 
coefficients $\mu _{1,2}(y)$ and $\rho (y)$ are $T$-periodic strictly
positive piecewise continuous functions of $y\in
\mathbb{R}
$, and $k,~\omega ~$are two real parameters (unless otherwise specified).
The functions $u(y)$ and $\mu _{1}(y)u^{\prime }(y)$ are assumed absolutely
continuous. They satisfy the quasi-periodic boundary conditions
\begin{equation}
u(T)=\mathrm{e}^{iKT}u(0),\quad \mu _{1}(T)u^{\prime }(T)=\mathrm{e}%
^{iKT}\,\mu _{1}(0)u^{\prime }(0)  \label{2}
\end{equation}%
with the Floquet parameter $K\in
\mathbb{C}
$, which by periodicity of $\mathrm{e}^{iKT}$ may be defined on the strip $%
\mathop{\rm Re}KT\in \left[ -\pi ,\pi \right] $ called the Brillouin zone.
Note that Eq.\ (\ref{1}) admits equivalent representations obtained by 
changing the function and/or variable. For instance, replacing the variable $%
y\Rightarrow \widetilde{y}=\int_{0}^{y}\mu _{1}^{-1}\left( \varsigma \right)
\mathrm{d}\varsigma $ recasts (\ref{1}) in the form of a weighted Schr\"{o}%
dinger equation
\begin{equation}
u^{\prime \prime }\left( \widetilde{y}\right) +\omega ^{2}Z^{2}u\left(
\widetilde{y}\right) =0,\ \mathrm{\ with}\ \omega ^{2}Z^{2}=\left( \omega
^{2}-\mu _{2}k^{2}/\rho \right) Z_{0}^{2},\ Z_{0}^{2}=\rho \mu _{1}.
\label{3.2}
\end{equation}%
Note that this transformation does not require reinforcing the above-imposed
condition of piecewise continuity of $\mu _{1}(y)$. The coefficients $Z$ and
$Z_{0\text{ }}$($Z=Z_{0}$ at $k=0$) have the physical meaning of,
respectively, impedance and normal impedance that we will find useful for
interpretations.

There exists a comprehensive spectral theory describing the eigenvalues $%
\omega _{n}^{2}$ ($n\in \mathbb{N}$) of (\ref{1}), (\ref{2}) as functions of $K$ at fixed $k,$ e.g. 
\cite{Brillouin,Krein,MW,RS,Marchenko,Allaire05}.
From this perspective, the spectrum for real $K\in
\mathbb{R}
$ is represented by the Floquet branches $\omega _{n}(K)$ on the $(\omega
,K) $-plane. Each branch spans a finite range on the $\omega $-axis, called
a passband, with a corresponding bounded solution $u_{n}(y)$. Separating
them are the ranges of $\omega $, called stopbands, where $\omega \in
\mathbb{R}
$ and $KT\in \pi
\mathbb{Z}
+i\left(
\mathbb{R}
\backslash 0\right) .$ Properties of the functional dependence of $\omega
_{n}(K)$ at fixed $k$ can be described by various analytical means. One of
the key ingredients of this theory is the so-called Lyapunov real-valued
function $\Delta (\omega ^{2})$ defined as the half trace of the monodromy
matrix (the propagator over a period). By this definition, $\Delta (\omega
^{2})=\cos KT$ determines the passbands and stopbands as the ranges $%
\left\vert \Delta (\omega ^{2})\right\vert \leq 1$ and $\left\vert \Delta
(\omega ^{2})\right\vert >1$, respectively.

The present work is concerned with the more general framework in which the
parameter $k$ is considered as an independent variable on top of $\omega $
and $K$. Keeping $\omega ^{2}$ as an eigenvalue of Eqs.\ (\ref{1})-(\ref{2})
now implies its dependence on two parameters: $\omega _{n}=\omega _{n}(K,k)$%
. For $K$ real, $\omega _{n}(K,k)$ is a multisheet surface whose sheets
projected on the $\left( \omega ,k\right) $-plane span the passband areas
bounded by the cutoff lines ($\left\vert \Delta \right\vert =1$) and
separated by the stopband areas. Cutting this surface by the planes $k=const$
and $\omega =const$ produces the Floquet branches and the isofrequency
(a.k.a. slowness) curves, respectively. Clearly, such perspective is
considerably richer than the one restricted to the Floquet curves at fixed $%
k.$ It is also important to note that the present study differs from the
two-parameter Sturm-Liouville problem with Dirichlet, Neumann and Robin
boundary conditions, which has been studied elsewhere, see e.g. \cite{BV,SPK}%
.

The structure and main results of the paper are as follows. Section 2
introduces complementary approaches based on differential operators $%
\mathcal{A}_{K}(k),$ $\mathcal{B}_{K}(\omega )$ defined by (\ref{1}), (\ref%
{2}) and on the matricant $\mathbf{M}\left( y,y_{0}\right) $ of the
equivalent differential system. The operators $\mathcal{A}_{K}(k),$ $%
\mathcal{B}_{K}(\omega )$ are self-adjoint and have a complete orthogonal
system of joint eigenfunctions, as shown in Appendix A1 by explicit
construction of their resolvent operators. The eigenvalues $\omega _{n}^{2}$
and $k_{n}^{2}$ of $\mathcal{A}_{K}(k)$ and $\mathcal{B}_{K}(\omega )$ are
then linked to the monodromy matrix $\mathbf{M}(T,0)$ with eigenvalues $%
\mathrm{e}^{\pm iK}$ via the generalized (depending on two parameters)
Lyapunov function $\Delta (\omega ^{2},k^{2})\equiv \frac{1}{2}\mathrm{trace}%
\mathbf{M}(T,0)=\cos KT$. Section 3 describes this function in some detail$.$
It is shown in \S 3.1 that $\Delta (\omega ^{2},k^{2})$ inside the passbands
$\left\vert \Delta \right\vert <1$ has non-zero first derivatives in both $%
\omega ^{2}$ and $k^{2},$ and that $\Delta (\omega ^{2})$ for fixed $k^{2}$
and $\Delta (k^{2})$ at fixed $\omega ^{2}$ each satisfies Laguerre's
theorem (by virtue of the estimates of $\Delta (\omega ^{2},k^{2})$ given in
Appendix A2). These two fundamental facts explain the regular structure of
the passband/stopband spectrum on the $(\omega ,k)$-plane. The WKB approach
\cite{Heading} is used in \S 3.2 to provide an insight into the asymptotic
behaviour of stopbands for continuous and piecewise continuous periodic
coefficients. Zero-width stopbands\textit{\ }(ZWS) are introduced and
analyzed in \S 3.3. Generalizing the concept of\textit{\ }degenerate gaps of
a one-parameter spectrum (e.g. \cite{MO,Ko,GPB}), ZWS are intersections of the analytical
cutoff curves  $\left\vert \Delta \right\vert =1$ with the $(\omega ,k)$%
-plane. It is shown that ZWS may or may not exist for an arbitrary periodic
profile of $\rho (y)$ and $\mu _{1,2}(y)$, are likely to exist for any profile
that is even about the period midpoint, and always exist for a periodically
bilayered structure. In the model cases, ZWS may also form infinite lines on
the $(\omega ,k)$-plane. Closed-form expressions for the partial derivatives
of $\Delta (\omega ^{2},k^{2})$ are obtained in \S 3.4. The derivative of
any order is a multiple integral of the product of, specifically, right
off-diagonal elements $M_{2}$ of the matricant $\mathbf{M}$ taken at
different points $y$ within the period and weighted by $\rho (y)$ and/or $%
\mu _{2}(y).$ An alternative representation is derived for the first-order
derivatives of $\Delta (\omega ^{2},k^{2})$ within the passbands by using
the eigenfunctions of $\mathcal{A}_{K}(k)$ and $\mathcal{B}_{K}(\omega )$.
The two equivalent formulas obtained for the first derivatives of $\Delta
(\omega ^{2},k^{2})$ provide an explicit meaning to their sign-definiteness
and offer useful complementary insight. In particular, it reveals some
interesting attributes of the function $M_{2}(y+1,y),$ whose zeros $(\omega
,k)$ are $y$-dependent solutions of the Dirichlet problem on $[y,y+T],$ see
\S 3.5. The properties of the Lyapunov function $\Delta (\omega
^{2},k^{2})\left( =\cos KT\right) $ and the expressions for its derivatives
established in Section 3 are then used in Section 4 to analyze principal
cuts of the dispersion surface $\omega _{n}(K,k).$ In \S 4.1, dependence $%
\omega (k)$ for fixed $K$ is studied. It is shown that if $K$ is real then
the curves $\omega _{n}(k)$ are monotonic (this may not be so for complex $K$%
) and they tend to the same linear asymptote $k\min_{y\in \left[ 0,T\right] }%
\left[ \mu _{2}(y)/\rho (y)\right] $ which is independent of $n.$ In \S 4.2,
the dependence $\omega (K)$ at fixed $k$ is discussed. For real $K$, the
first non-zero derivative of Floquet branches $\omega _{n}(K)$ is provided
(it is a first derivative inside the passbands and a second one at the
cutoffs); for the stopbands, the condition on $\omega $ realizing maximum of
$\left\vert \mathop{\rm Im}K\left( \omega \right) \right\vert $ is
formulated. The real isofrequency curves $K(k)$ at fixed $\omega $ are
considered in \S \S 4.3 and 4.4. Particular attention is given to the closed
isofrequency curve arising for $\omega $ less than the first cutoff $\omega
_{1}\left( \pi T^{-1},0\right) .$ It is proved that, whatever the distortion of
its shape due to unidirectional periodicity may be, this isofrequency curve
is always convex and hence low-frequency caustics of the group velocity $%
\mathbf{\nabla }\omega $ are impossible. Finally, useful bounds on the first
eigenvalue $\omega _{1}(K,k)$ for $KT\in \left[ -\pi ,\pi \right] $ and any $%
k$ are provided in Appendix A3.

Without loss of generality, in the following we take $T=1;$ more precisely,
this implies the redefinitions $y\Rightarrow y/T\equiv y,$ $\omega
\Rightarrow \omega T\equiv \omega ,$ $k\Rightarrow kT\equiv k$ and $%
K\Rightarrow KT\equiv K$ so that the variables $y$ and $\omega ,$~$k,$~$K$
are hereafter non-dimensional. We also assume throughout that $T=1$ is a
\textit{minimal} possible period.

\section{ Eigenvalue problem, monodromy matrix and Lyapunov function}

Equation (\ref{1}) with the conditions (\ref{2}) can be considered in either
of the equivalent forms
\begin{equation}
\mathcal{A}_{K}u=\omega ^{2}u,\quad \mathcal{B}_{K}u=k^{2}u,\quad u\in D_{K}
\label{4}
\end{equation}%
with the operators $\mathcal{A}_{K}\equiv \mathcal{A}_{K}(k)$ and $\mathcal{B%
}_{K}\equiv \mathcal{B}_{K}(\omega )$
\begin{equation}
\mathcal{A}_{K}u=-\frac{1}{\rho }\left( \mu _{1}u^{\prime }\right) ^{\prime
}+k^{2}\frac{\mu _{2}}{\rho }u,\quad \mathcal{B}_{K}u=\frac{1}{\mu _{2}}%
\left( \mu _{1}u^{\prime }\right) ^{\prime }+\omega ^{2}\frac{\rho }{\mu _{2}%
}u.  \label{5}
\end{equation}%
Their common domain is
\begin{equation}
\begin{array}{c}
D_{K}=\left\{ u\in D:\ \mathbf{\eta }\left( 1\right) =\mathrm{e}^{iK}\mathbf{%
\eta }(0)\right\} , \\
D=\left\{ u\in AC\left[ 0,1\right] ,\ \mu _{1}u^{\prime }\in AC\left[ 0,1%
\right] \right\} ,%
\end{array}%
\ \ \ \mathbf{\eta }(y)=%
\begin{pmatrix}
u(y) \\
i\mu _{1}(y)u^{\prime }(y)%
\end{pmatrix}%
,  \label{6}
\end{equation}%
where $K\in
\mathbb{C}
$ and $AC[0,1]$ is the space of all absolutely continuous functions from $%
[0,1]$ to $%
\mathbb{C}
$ (note that using "$i$" in the definition of $\mathbf{\eta }$ and hence in (%
\ref{10})$_{2}$ is a conventional option which is useful for a compact form
of (\ref{13})$_{1}$ and similar identities). Let $\left( \cdot ,\cdot
\right) _{\rho ,~\mu _{2}}$ and $\left\Vert \cdot \right\Vert _{\rho ,~\mu
_{2}}$ be a standard inner product and norm in the Hilbert space $\mathcal{H}%
_{\rho ,~\mu _{2}}=L_{\rho ,~\mu _{2}}^{2}\left( 0,1\right) $ of functions
with quadratically summable measure $\rho (y)\mathrm{d}y$ and $\mu _{2}(y)%
\mathrm{d}y,$ respectively; so that
\begin{equation}
\begin{aligned} \left( u,v\right) _{\rho }& =\int_{0}^{1}\rho (y)u(y)v^{\ast
}(y)\mathrm{d}y,\quad \left\Vert u\right\Vert _{\rho }^{2}=\left( u,u\right)
_{\rho }, \\ \left( u,v\right) _{\mu _{2}}& =\int_{0}^{1}\mu
_{2}(y)u(y)v^{\ast }(y)\mathrm{d}y,\quad \left\Vert u\right\Vert _{\mu
_{2}}^{2}=\left( u,u\right) _{\mu _{2}}, \end{aligned}  \label{7}
\end{equation}%
where $^{\ast }$ means complex conjugation.

The operator (\ref{1}) on $L^{2}\left(
\mathbb{R}
\right) $ with eigenvalues $\omega ^{2}$ (or $k^{2}$) can be represented as
a direct integral decomposition $\oplus _{K\in \left[ 0,2\pi \right] }%
\mathcal{A}_{K}$ (or $\oplus _{K\in \left[ 0,2\pi \right] }\mathcal{B}_{K}$)
\cite{RS}. Therefore the spectrum of the operator (\ref{1}) is a union of
all eigenvalues of $\mathcal{A}_{K}$ (or $\mathcal{B}_{K}$) for $K\in \left[
0,2\pi \right] $ and hence for all $K\in
\mathbb{R}
$ since $\mathcal{A}_{K}=\mathcal{A}_{K+2\pi },$ $\mathcal{B}_{K}=\mathcal{B}%
_{K+2\pi }.$ The operators $\mathcal{A}_{K}$ and $\mathcal{B}_{K}$ are
symmetric if $K\in
\mathbb{R}
$, i.e. $\left( \mathcal{A}_{K}u,v\right) _{\rho }=\left( u,\mathcal{A}%
_{K}v\right) _{\rho },$\ $\left( \mathcal{B}_{K}u,v\right) _{\mu
_{2}}=\left( u,\mathcal{B}_{K}v\right) _{\mu _{2}}\ $for $u,v\in D_{K}$, and
they both have compact and self-adjoint resolvents that satisfy the
Hilbert-Schmidt theorem (see Appendix A1). Therefore $\mathcal{A}_{K}$ and $%
\mathcal{B}_{K}$ are self-adjoint with purely discrete spectra $\sigma
\left( \mathcal{A}_{K}\right) $ and $\sigma \left( \mathcal{B}_{K}\right) $
containing an infinite number of real eigenvalues $\omega _{n}^{2}\left(
K,k\right) $ and $k_{n}^{2}\left( K,\omega \right) $ ($n\in
\mathbb{N}
$), and corresponding eigenfunctions $u_{n}$($\equiv u_{n,\mathcal{A}}$ and $%
u_{n,\mathcal{B}}$) forming a complete orthogonal system in the spaces $%
\mathcal{H}_{\rho }$ and $\mathcal{H}_{\mu _{2}},$ respectively. The
operator $\mathcal{A}_{K}$ is positive for any $k\in
\mathbb{R}
$ $\ $(i.e. for any $k^{2}\geq 0$),
\begin{equation}
\left( \mathcal{A}_{K}u,u\right) _{\rho }\geq 0\ \ \ \left( >0\ \mathrm{at}\
k\neq 0\right) ,  \label{9}
\end{equation}%
so its spectrum $\sigma \left( \mathcal{A}_{K}\right) $ consists of
non-negative eigenvalues $\omega _{n}^{2}(K,k)$ (strictly positive at $k\neq
0$), which are hereafter numbered in increasing order $\omega _{1}\leq
\omega _{2}\leq \ldots $ By contrast, $\mathcal{B}_{K}$ is not sign-definite
and hence its spectrum $\sigma \left( \mathcal{B}_{K}\right) $ includes both
positive and negative eigenvalues $k_{n}^{2}\left( K,\omega \right) $. Note
that real eigenvalues of $\mathcal{A}_{K}$\textit{\ }and\textit{\ }$\mathcal{%
B}_{K}$\textit{\ }are also admitted at $\mathop{\rm Im}K\neq 0$\textit{\ }%
(see Definition \ref{14.1}(c) below).

Equation (\ref{1}) can be recast as
\begin{equation}
\mathbf{\eta }^{\prime }(y)=\mathbf{Q}(y)\mathbf{\eta }(y)\ \mathrm{with}\
\mathbf{Q}(y)=i%
\begin{pmatrix}
0 & -\mu _{1}^{-1} \\
\mu _{2}k^{2}-\rho \omega ^{2} & 0%
\end{pmatrix}
\label{10}
\end{equation}%
for $\mathbf{\eta }(y)$ introduced in (\ref{6})$_{2}$. Given an initial
condition $\mathbf{\eta }\left( y_{0}\right) $, Eq.\ (\ref{10})$_{1}$ has a
unique solution%
\begin{equation}
\mathbf{\eta }(y)=\mathbf{M}\left( y,y_{0}\right) \mathbf{\eta }\left(
y_{0}\right)  \label{11}
\end{equation}%
defined through the propagator matrix, or matricant,
\begin{align}
\mathbf{M}\left( y,y_{0}\right) & \equiv
\begin{pmatrix}
M_{1}\left( y,y_{0}\right) & M_{2}\left( y,y_{0}\right) \\
M_{3}\left( y,y_{0}\right) & M_{4}\left( y,y_{0}\right)%
\end{pmatrix}%
=\widehat{\int }_{y_{0}}^{y}\left[ \mathbf{I}+\mathbf{Q}\left( \varsigma
\right) \mathrm{d}\varsigma \right]  \notag \\
& =\mathbf{I}+\int_{y_{0}}^{y}\mathbf{Q}\left( \varsigma _{1}\right) \mathrm{%
d}\varsigma _{1}+\int_{y_{0}}^{y}\mathbf{Q}\left( \varsigma _{1}\right)
\mathrm{d}\varsigma _{1}\int_{y_{0}}^{\varsigma _{1}}\mathbf{Q}\left(
\varsigma _{2}\right) \mathrm{d}\varsigma _{2}+\ldots ,  \label{12}
\end{align}%
where $\widehat{\int }$ is the multiplicative integral evaluated by the
Peano series \cite{Pease} and $\mathbf{I}$ is the 2$\times $2 identity matrix.
Note that $\det \mathbf{M}\left( y,y_{0}\right) =1$ due to $\mathrm{tr}%
\mathbf{Q}=0,$ where $\mathrm{tr}$ means the trace. By (\ref{10}) $\mathbf{Q}%
=-\mathbf{TQ}^{+}\mathbf{T}$ for $\omega ^{2},~k^{2}\in
\mathbb{R}
$ and so
\begin{equation}
\mathbf{M}^{-1}\left( y,y_{0}\right) =\mathbf{TM}^{+}\left( y,y_{0}\right)
\mathbf{T}\ \Rightarrow \ \mathop{\rm Im}M_{1,4}\left( y,y_{0}\right) =0,\ %
\mathop{\rm Re}M_{2,3}\left( y,y_{0}\right) =0,  \label{13}
\end{equation}%
where $^{+}$ denotes Hermitian transpose and $\mathbf{T}$ is the 2$\times $2
matrix with zero diagonal and unit off-diagonal elements. If $\mathbf{Q}%
\left( y\right) $ is also even about the midpoint of the interval $\left[
y_{0},y\right] $ then%
\begin{equation}
\mathbf{M}\left( y,y_{0}\right) =\mathbf{TM}^{T}\left( y,y_{0}\right)
\mathbf{T}\ \Rightarrow \ M_{1}\left( y,y_{0}\right) =M_{4}\left(
y,y_{0}\right) ,  \label{13a}
\end{equation}%
where $^{T}$ denotes transpose. The properties (\ref{13})$_{1}$ and (\ref%
{13a})$_{1}$ are actually valid for matrices $\mathbf{Q}$ and $\mathbf{M}$
of arbitrary $n\times n$ size (see \cite{SPK} for details), while (\ref{13}$%
_{2}$) and (\ref{13a})$_{2}$ are attributes of the 2$\times $2 case which
admits easy direct proofs (e.g. (\ref{13})$_{2}$ is evident from the
definition (\ref{6})$_{2}$ of $\mathbf{\eta }$ with a real scalar $u$).

Assume a periodic $\mathbf{Q}(y) $ so that $\mathbf{Q}(y) =\mathbf{Q}\left(
y+1\right) $ and hence $\mathbf{M}\left( y,y_{0}\right) =\mathbf{M}\left(
y+1,y_{0}+1\right) $. The propagator $\mathbf{M}\left( y_{0}+1,y_{0}\right) $
over a period $\left[ y_{0},y_{0}+1\right] $ is called the \emph{monodromy}
matrix. For any $y_{0}\equiv y,$ denote its elements as
\begin{equation}
\mathbf{M}( y+1,y) =
\begin{pmatrix}
m_{1}(y) & im_{2}(y) \\
im_{3}(y) & m_{4}(y)%
\end{pmatrix}
,\ \
\begin{array}{c}
m_{1,4}(y) =M_{1,4}( y+1,y) , \\
im_{2,3}(y) =M_{2,3}( y+1,y) ,%
\end{array}
\label{13.1}
\end{equation}%
where $\mathop{\rm Im}m_{j}(y) =0,\ j=1..4,$ for $\omega ^{2},~k^{2}\in
\mathbb{R}
$ by (\ref{13})$_{2}$. The assumed periodicity with use of the chain rule
implies the identity
\begin{equation}
\mathbf{M}( y+1,y) =\mathbf{M}\left( y+1,1\right) \mathbf{M}(1,0) \mathbf{M}%
( 0,y) =\mathbf{M}( y,0) \mathbf{M}(1,0) \mathbf{M}^{-1}(y,0) .  \label{13.2}
\end{equation}

\begin{remark}
\label{13.3} The trace and eigenvalues of $\mathbf{M}\left(y+1,y\right) $
are independent of $y$ by virtue of (\ref{13.2}).
\end{remark}

\noindent Hereafter, unless otherwise specified, we set $y_{0}=0$ and define
the monodromy matrix as $\mathbf{M}(1,0)$ with respect to the period $\left[
0,1\right] $ (as in (\ref{6}), (\ref{7})).

Bearing in mind $\det \mathbf{M}=1,$ denote the eigenvalues of $\mathbf{M}%
(1,0) $ by $q$ and $q^{-1}.$ Introduce the generalized Lyapunov function
\begin{equation}
\Delta (\omega ^{2},k^{2})\equiv \frac{1}{2}\mathrm{tr}\mathbf{M}(1,0) =%
\frac{1}{2}\left( q+q^{-1}\right) ,  \label{14.0}
\end{equation}%
which is analytic in $\omega ^{2},k^{2}$ by (\ref{10})$_{2}$, (\ref{12}) and
real for $\omega ^{2},~k^{2}\in
\mathbb{R}
$ by (\ref{13})$_{2}$. As noted above, the function $\Delta (\omega
^{2},k^{2})$ is independent of the interval on which the unit period is
defined. It is also invariant for any similarity equivalent formulation of
the system matrix $\widetilde{\mathbf{Q}}(y) =\mathbf{C}^{-1}\mathbf{Q}(y)
\mathbf{C}$ because $\mathrm{tr}\widetilde{\mathbf{M}}=\mathrm{tr}\left(
\mathbf{C}^{-1}\mathbf{MC}\right) =\mathrm{tr}\mathbf{M}$, leaving $\Delta
(\omega ^{2},k^{2})$ unchanged. \

\begin{proposition}
\label{14.01}For any complex numbers $k,\ \omega ,\ K$, the following
statements are equivalent: (i) $\omega ^{2}$ is an eigenvalue of the
operator $\mathcal{A}_{K}(k) ;$ (ii) $k^{2}$ is an eigenvalue of the
operator $\mathcal{B}_{K}( \omega ) ;$\ (iii) $k,\ \omega $ and $K$ are
connected by the equality%
\begin{equation}
\Delta (\omega ^{2},k^{2})-\cos K=0.  \label{14}
\end{equation}
\end{proposition}

\noindent \textit{Proof.} The link \textit{(i)}$\Rightarrow $\textit{(ii)}
follows from Eq.\ (\ref{4}). Consider \textit{(i),(ii)}$\Rightarrow $\textit{%
(iii).} According to \textit{(i)} or \textit{(ii),} $\omega ^{2}$ or $k^{2}$
is an eigenvalue of, respectively, $\mathcal{A}_{K}(k) $ or $\mathcal{B}%
_{K}( \omega ) $. Then there exists $u(y)\in D_{K}$ that satisfies (\ref{4})
hence (\ref{1}), and consequently the vector $\mathbf{\eta }(y),$ generated
by $u(y)$ according to (\ref{6})$_{2}$, is a solution of Eq.\ (\ref{10}).
So, by (\ref{11}), $\mathbf{\eta }\left( 1\right) =\mathbf{M}(1,0) \mathbf{%
\eta }(0) .$ On the other hand, as indicated in (\ref{6})$_{1}$, $u(y)\in
D_{K}$ implies that $\mathbf{\eta }\left( 1\right) =\mathrm{e}^{iK}\mathbf{%
\eta }\left( 0\right) $. Hence $\mathrm{e}^{iK}$ is an eigenvalue $q$ of $%
\mathbf{M}(1,0) $, and the function $\Delta $ defined by (\ref{14.0})
satisfies (\ref{14}), that is \textit{(iii)}. Now consider \textit{(iii)}$%
\Rightarrow $\textit{(i),(ii)}. From (\ref{14}) and the definition (\ref%
{14.0}), the eigenvalue $q$ of $\mathbf{M}(1,0) $ is $q=\mathrm{e}^{iK}$,
and corresponding eigenvector $\mathbf{w}$ exists such that $\mathbf{M}(1,0)%
\mathbf{w}=\mathrm{e}^{iK}\mathbf{w}.$ Let $u(y)$ be the first component of
the solution $\mathbf{\eta }(y) =\mathbf{M}(y,0) \mathbf{w}$ of Eq.\ (\ref%
{10}) with the initial condition $\mathbf{\eta }(0) =\mathbf{w}$. From the
above, $u(y)$ belongs to $D_{K}$ and satisfies Eq.\ (\ref{4}), which implies
\textit{(i),(ii)}. $\blacksquare $

\begin{corollary}
\label{14.02}Each eigenfunction $u$ of $\mathcal{A}_{K}$ and $\mathcal{B}%
_{K} $ is equal to the first component of the vector $\mathbf{\eta }(y) =%
\mathbf{M}(y,0) \mathbf{w,}$ where $\mathbf{w}$ is the eigenvector of $%
\mathbf{M}(1,0) $ corresponding to the eigenvalue $q=\mathrm{e}^{iK}.$
\end{corollary}

\begin{definition}
\label{14.1} Passband areas, cutoffs and stopband areas are defined for $%
\omega ^{2},~k^{2}\in \mathbb{R} $ (and hence real $\Delta ( \omega
^{2},k^{2} ) $) as follows:
\begin{equation}
\left( \omega ,~k\right) :
\begin{cases}
\left\vert \Delta \right\vert \leq1 \ \ \, \left( \Leftrightarrow K\in
\mathbb{R} \right) & \text{passbands}, \\
\Delta =\pm 1 \ \ \left( \Leftrightarrow K\in \pi\mathbb{Z}\right) & \text{%
cutoffs} , \\
\left\vert \Delta \right\vert >1 \ \ \, \left( \Leftrightarrow K\in \pi
\mathbb{Z} +i\left( \mathbb{R} \backslash 0\right) \right) & \text{stopbands}
.%
\end{cases}
\notag
\end{equation}
\end{definition}

Before discussing general properties of the Lyapunov function $\Delta
(\omega ^{2},k^{2}),$ it is expedient to mention its explicit properties at $%
\omega =0$ and/or $k=0$. Obviously $\partial \Delta /\partial \omega =0$ at $%
\omega =0$ and $\partial \Delta /\partial k=0$ at $k=0.$ By (\ref{10})$_{2}$, (\ref{12}) and (\ref{14.0}),%
\begin{equation}
\begin{aligned} \Delta (\omega ^{2},k^{2}) &=1+\tfrac{1}{2}\left\langle \mu
_{1}^{-1}\right\rangle \left( \left\langle \mu _{2}\right\rangle
k^{2}-\left\langle \rho \right\rangle \omega ^{2}\right) +O\big( ( \omega
^{2}+k^{2})^{2}\big) \ \mathrm{with}\ \left\langle \cdot \right\rangle
\equiv \int_{0}^{1}\left( \cdot \right) \mathrm{d}y; \\ \partial \Delta
/\partial ( \omega ^{2} ) &=-\tfrac{1}{2} \left\langle \rho \right\rangle
\left\langle \mu _{1}^{-1}\right\rangle ,\ \partial \Delta /\partial \left(
k^{2}\right) =\tfrac{1}{2}\left\langle \mu _{1}^{-1}\right\rangle
\left\langle \mu _{2}\right\rangle \ \mathrm{at}\ \omega =0,\,k=0,
\end{aligned}  \label{19}
\end{equation}%
\noindent where the identity $\int_{0}^{1}\mathrm{d}\varsigma
\int_{0}^{\varsigma _{1}}\left[ f_{1}\left( \varsigma \right) f_{2}\left(
\varsigma _{1}\right) +f_{2}\left( \varsigma \right) f_{1}\left( \varsigma
_{1}\right) \right] \mathrm{d}\varsigma _{1}=\left\langle f_{1}\right\rangle
\left\langle f_{2}\right\rangle $ was used in (\ref{19})$_{1}$. Note that $%
\Delta (0,k^{2})>1$ for$\mathrm{\ }k^{2}>0$ and $\left[ \partial \Delta
/\partial ( \omega ^{2} ) \right] _{\omega =0}<0$ for $k^{2}\geq 0, $
whereas the bounds of $\Delta (\omega ^{2},0)$ and the sign of $\left[
\partial \Delta /\partial \left( k^{2}\right) \right] _{k=0}$ are not fixed
for $\omega ^{2}>0.$ Also note the explicit non-semisimple form of the
matrix
\begin{equation}
\mathbf{M}(y,0) =
\begin{pmatrix}
1 & -i\int_{0}^{y}\mu _{1}^{-1}\left( \varsigma \right) \mathrm{d}\varsigma
\\
0 & 1%
\end{pmatrix}
\ \mathrm{at}\ \omega =0,\,k=0.  \label{19.1}
\end{equation}

\section{Properties of the Lyapunov function $\Delta (\protect\omega %
^{2},k^{2})$}

\subsection{Formation of the passband/stopband spectrum}

\noindent We proceed with  some observations on the analytical properties of
the function $\Delta (\omega ^{2},k^{2})$ that underlie the alternating
structure of the passbands and stopbands.

\begin{lemma}
\label{19.10}If $\omega \notin
\mathbb{R}
$ or $k^{2}\notin
\mathbb{R}
$ then $\Delta \notin \left[ -1,1\right] .$
\end{lemma}

\noindent \textit{Proof.} If $\Delta \in \left[ -1,1\right] $ then according
to Proposition \ref{14.01} the identity (\ref{14}) holds for $K\in
\mathbb{R}
$ and hence $\omega ^{2}$ or $k^{2}$ is an eigenvalue of $\mathcal{A}_{K}(k)
$ or $\mathcal{B}_{K}( \omega ) $, respectively. It was shown (see (\ref{11}%
) and below) that the eigenvalues of $\mathcal{A}_{K}(k) $ are positive and
the eigenvalues of $\mathcal{B}_{K}( \omega ) $ are real. $\blacksquare $

\begin{proposition}
\label{19.00}The derivatives $\partial \Delta /\partial ( \omega ^{2}) $ and
$\partial \Delta /\partial \left( k^{2}\right) $ do not vanish within an
open passband interval $\Delta (\omega ^{2},k^{2})\in \left( -1,1\right) .$
\end{proposition}

\noindent \textit{Proof.} By Lemma \ref{19.10}, if $\Delta \in \left(
-1,1\right) $ then $\omega ^{2},k^{2}\in
\mathbb{R}
.$ Suppose that $\partial \Delta /\partial ( \omega ^{2} ) =0$ for some real
value $\omega ^{2}.$ Then, because $\Delta (\omega ^{2})$ ($\equiv \Delta
(\omega ^{2},k^{2})$ at fixed $k$) is an analytic function, there exists
complex $\widetilde{\omega }^{2}$ in the vicinity of $\omega ^{2}$ for which
$\Delta (\widetilde{\omega }^{2})\in \left( -1,1\right) $ . This contradicts
Lemma \ref{19.10}, and hence $\partial \Delta /\partial \left( \omega
^{2}\right) \neq 0.$ The same reasoning proves that $\partial \Delta
/\partial \left( k^{2}\right) \neq 0$. Consequently, Eq.\ (\ref{14}) at
fixed $\omega ^{2}>0$ (or fixed real $k^{2}$) has only real and simple roots
$k_{n}^{2}$ (or $\omega _{n}^{2}$) if $\cos K\in \left( -1,1\right) .$ $%
\blacksquare $

\noindent Proposition \ref{19.00} plays a pivotal role in explaining the
origin of the Floquet stopbands by the following simple reasoning. Consider $%
\rho (y) ,$ $\mu _{1,2}(y) $ resulting from an arbitrary periodic
perturbation of some reference constant values $\rho _{0}$ and $\mu
_{01,02}, $ so that $\Delta (\omega ^{2},k^{2})$ is a perturbation of $%
\Delta _{0}(\omega ^{2},k^{2})=\cos K$ with $K^{2}=\frac{\rho _{0}}{\mu _{01}%
}\omega ^{2}-\frac{\mu _{02}}{\mu _{01}}k^{2}.$ Since the first derivatives
of $\Delta (\omega ^{2},k^{2})$ do not vanish within $\left( -1,1\right) ,$
the perturbed extreme values $\Delta _{0}=\pm 1$ must either remain equal to
$\pm 1$ or exceed the range $\left[ -1,1\right] ,$ thereby leading to
complex values $K\in \pi
\mathbb{Z}
+i\left(
\mathbb{R}
\backslash 0\right) $, i.e., to the stopbands.

\begin{proposition}
\label{19.20}For $\omega ^{2},k^{2}\in
\mathbb{R}
,$ the derivatives of any order $n\in
\mathbb{N}
$ of the functions $\Delta (\omega ^{2})$ and $\Delta (k^{2})$ ($\equiv
\Delta (\omega ^{2},k^{2})$ at fixed $k$ and fixed $\omega $, respectively)
have only real and simple zeros, each lying between consecutive zeros of the
$\left( n-1\right) $th derivative of the same function. In particular, the
first derivatives of $\Delta (\omega ^{2})$ and $\Delta (k^{2})$ have a
single and simple zero between consecutive zeros of $\Delta (\omega
^{2},k^{2})$ and do not vanish elsewhere.
\end{proposition}

\noindent \textit{Proof.} It is shown in Appendix A2 that the functions $%
\Delta (\omega ^{2})$ and $\Delta (k^{2})$ are entire functions of order of
growth
$\frac12$%
. Their zeros are the eigenvalues of the operators $\mathcal{A}_{\pi /2}(k) $
and $\mathcal{B}_{\pi /2}( \omega ) $, and are therefore real and simple.
Hence both functions satisfy the conditions of Laguerre's theorem (e.g. \cite{Titchmarsh}), implying that the derivatives of $\Delta (\omega ^{2})$ and of $\Delta
(k^{2})$ are also entire functions with order of growth
$\frac12$
and they have the desired properties. $\blacksquare $

\begin{figure}[H] \centering
 \parbox[b]{0.49\textwidth}{\centering
 \includegraphics{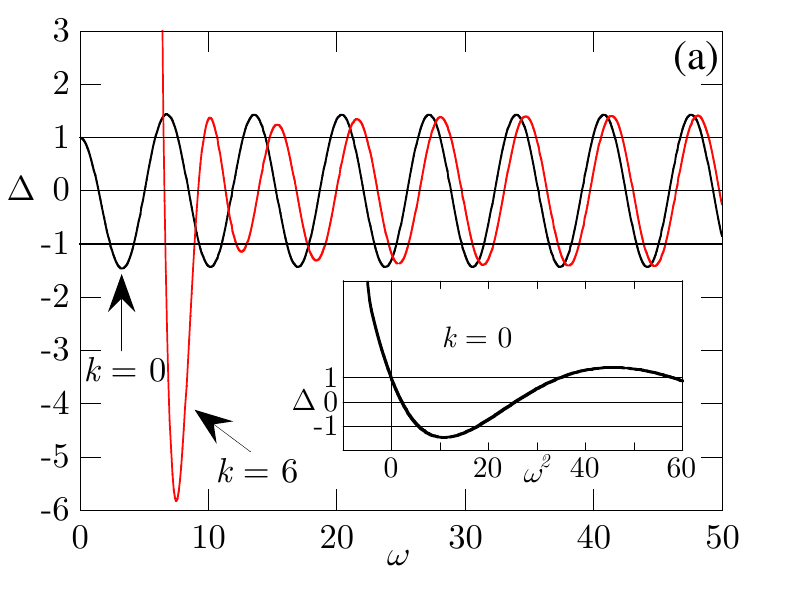}
} \hfil \hfil
\begin{minipage}[b]{0.49\textwidth}
\centering 
\includegraphics{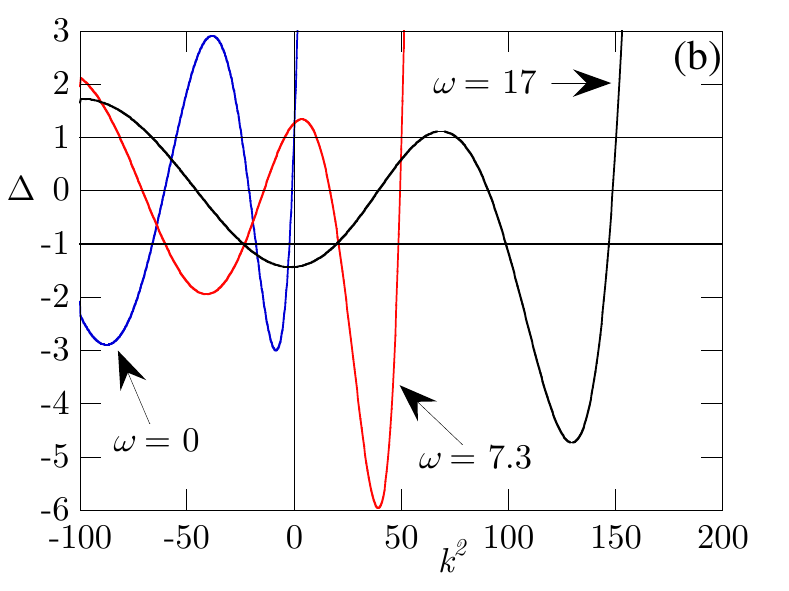}
\end{minipage}
\caption{Generalized Lyapunov function $\Delta (\protect\omega ^{2},k^{2})$%
\textbf{\ }for the profile (\protect\ref{0}): (a) $\Delta (\protect\omega %
)\left( =\Delta (-\protect\omega )\right) $ at different fixed
values of $k$
(a fragment of $\Delta (\protect\omega ^{2})$ at $k=0$ for $\protect\omega %
^{2}\gtrless 0$ is shown in the inset); (b) $\Delta (k^{2})$ at
different fixed values $\protect\omega $.}\label{fig1}
\end{figure}

Propositions \ref{19.00} and \ref{19.20} define the basic form of the
function $\Delta (\omega ^{2},k^{2})$ at fixed $\omega $ or $k$. It is
exemplified in Fig.\ \ref{fig1} for a piecewise continuous profile of
material coefficients chosen as
\begin{equation}
\mu _{1}(y)=\mu _{2}(y)=\frac{1}{4}(1+3y)^{2}(2+y),\ \rho (y)=2+y\ \mathrm{%
for}\ y\in \lbrack 0,1]  \label{0}
\end{equation}%
(taking $\mu _{1,2}$ in GPa and $\rho $ in g/cm$^{3}$ implies $\omega
T\equiv \omega $ in MHz$\cdot $mm in this and subsequent figures). Note that
$\Delta (\omega ^{2})$ has an infinite number of zeros that are strictly
positive and move rightwards as $k$ increases, whereas $\Delta (k^{2})$ has
an infinite number of negative zeros at $\omega =0$ which move one by one on
the positive semi-axis $k^{2}>0$ as $\omega $ increases.

Since zeros of the first derivatives of $\Delta (\omega ^{2},k^{2})$ cannot
be points of inflection or zero-curvature by Proposition \ref{19.20}, we can
now refine the numbering of branches $\omega _{n}(K,k)=$ $\sqrt{\omega
_{n}^{2}(K,k)}$ $\left( \geq 0\right) $ in the passbands as follows:
\begin{equation}
\begin{aligned} & 0<\omega _{1}(K,k)<\omega _{2}(K,k)<\ldots \  &&
\mathrm{if}\ K\in \mathbb{R} ,\ K\notin \pi\mathbb{Z}; \\ & 0\leq \omega
_{1}(0,k)<\omega _{2}(0,k)\leq \omega _{3}(0,k)<\omega _{4}(0,k)\leq \ldots
\  &&\mathrm{if}\ K\in 2\pi \mathbb{Z} ; \\ & 0<\omega _{1}(\pi ,k)\leq
\omega _{2}(\pi ,k)<\omega _{3}(\pi ,k)\leq \omega _{4}(\pi ,k)<\ldots \
&&\mathrm{if}\ K\in \pi +2\pi \mathbb{Z}. \end{aligned}  \label{19.3}
\end{equation}%
With reference to (\ref{19}) and Proposition \ref{19.00}, the sign of first
derivatives of $\Delta (\omega ^{2},k^{2})$ along $\omega _{n}(K,k)$ in the $%
n$th open passband $\left\vert \Delta \right\vert <1$ (see ({(\ref{19.3}$%
_{1} $})) is
\begin{equation}
\mathrm{sgn}\left[ \partial \Delta /\partial (\omega ^{2})\right] =-\mathrm{%
sgn}\left[ \partial \Delta /\partial \left( k^{2}\right) \right] =\left(
-1\right) ^{n}.  \label{19.2}
\end{equation}%
{The possibility of equality of two cutoffs (see (\ref{19.3})$_{2,3}$), i.e.
of a double root of the equation} $\Delta (\omega ^{2})=\pm 1$, implies a
zero-width stopband addressed in detail in \S 3.3.

For the future use, let us also mention some properties of the Dirichlet and
Neumann eigenvalues $\omega _{\mathrm{D},n}^{2}$ and $\omega _{\mathrm{N}%
,n}^{2}$ of (\ref{1}) satisfying the conditions $u\left( 0\right) =0,$ $%
u\left( 1\right) =0$ and $u^{\prime }\left( 0\right) =0,$ $u^{\prime }\left(
1\right) =0$, respectively. It is known that $\omega _{\mathrm{D},n}\ $and\ $%
\omega _{\mathrm{N},n}$ are simple zeros of the functions $M_{2}(1,0)$ and $%
M_{3}(1,0)$ of $\omega ,$ which occur once per each stopband complemented by
cutoffs (except the first stopband devoid of $\omega _{\mathrm{D},n}$). The
branches $\omega _{\mathrm{D},1}(k)<~\omega _{\mathrm{D},2}\left( k\right)
...$ and $\omega _{\mathrm{N},1}(k)<\omega _{\mathrm{N},2}(k)...$ are thus
related to the passband eigenvalues $\omega _{n}\left( K,k\right) $ of (\ref%
{19.3}) as
\begin{equation}
\omega _{\mathrm{D},2j}(k),\omega _{\mathrm{N},2j+1}(k)\in \lbrack \omega
_{2j}(0,k),\omega _{2j+1}(0,k)];\ \ \omega _{\mathrm{D},2j-1}(k),\omega _{%
\mathrm{N},2j}(k)\in \lbrack \omega _{2j-1}(\pi ,k),\omega _{2j}(\pi ,k)],
\label{19.4}
\end{equation}%
where $j\in
\mathbb{N}
$ and $\omega _{\mathrm{N},1}(k)\in \lbrack 0,\omega _{1}(0,k)]$. Recall
that the stopbands and cutoffs are invariant with respect to the choice of
the period interval $\left[ y_{0},y_{0}+1\right] \equiv \left[ 0,1\right] $
(see Remark \ref{13.3}); however, the branches $\omega _{\mathrm{D},n}(k)$%
~and\ $\omega _{\mathrm{N},n}(k)$ within this area certainly depend on the
choice of the point $y_{0}\equiv 0$. In other words, some fixed values $%
\omega ,$~$k$ realize the Dirichlet or Neumann conditions at the edges of $%
\left[ y_{0},y_{0}+1\right] $ iff $y_{0}$ is a zero of the function $%
M_{2}\left( y+1,y\right) \equiv im_{2}(y)$ or $M_{3}\left( y+1,y\right)
\equiv im_{3}(y)$, respectively (see \S 3.5 for further discussion).
According to (\ref{13a}), if $\mathbf{Q}(y)$ is an even function about the
midpoint of the period $\left[ y_{0},y_{0}+1\right] $ for some $y_{0}$, then
the Dirichlet and Neumann branches $\omega _{\mathrm{D},n}(k)$~and\ $\omega
_{\mathrm{N},n}(k)$ satisfying $m_{2}(y_{0})=0$ and $m_{3}(y_{0})=0$
coincide with the cutoff curves. We note the useful identity $m_{2}(y)m_{3}(y)>0$
for $\left\vert \Delta \right\vert <1$ which may be proved as follows: it
obviously holds for $\Delta =0$ due to $\det \mathbf{M}=1$, and hence for
any $\left\vert \Delta \right\vert <1$ due to the fact that $m_{2}(y)$ and $%
m_{3}(y)$ are strictly non-zero inside the passbands by (\ref{19.4}).

\subsection{WKB asymptotics of $\Delta $}

Some insight into the high-frequency spectrum in the case of continuous and
piecewise continuous periodicity can be gained from the WKB asymptotics \cite{Heading} of the Lyapunov function $\Delta (\omega ^{2},k^{2})$ at fixed $k$. To this
end recall the impedance $Z=Z_{0}\sqrt{1-\mu _{2}k^{2}/\rho \omega ^{2}}$
with $Z_{0}=\sqrt{\rho \mu _{1}}$ introduced in (\ref{3.2}). For any fixed $%
k,$ let $\omega ^{2}>k^{2}\max_{y\in \left[ 0,1\right] }\left( \mu _{2}/\rho
\right) $ so that $Z(y)$ is real (the so-called supersonic regime). Suppose
for brevity that the overall periodic profile of $Z(y)$ has at most one
point of discontinuity per period. If so, the zero-order WKB approximation $%
\Delta _{\mathrm{WKB}}^{(0)}$ of $\Delta $ takes an especially simple form
\begin{equation}
\Delta _{\mathrm{WKB}}^{(0)}=\frac{1}{2}\big(\left[ Z\right] ^{1/2}+\left[ Z%
\right] ^{-1/2}\big)\cos \big(\omega \int_{0}^{1}\mu _{1}^{-1}Z\mathrm{d}y%
\big),  \label{WKB}
\end{equation}%
where $\pm i\omega \mu _{1}^{-1}Z$ are the eigenvalues of the matrix $%
\mathbf{Q}$ defined in (\ref{10})$_{2}$ and $\left[ Z\right] =Z\left(
y_{d}^{-}\right) /Z\left( y_{d}^{+}\right) $ with $Z\left( y_{d}^{\pm
}\right) \equiv \lim_{\varepsilon \rightarrow 0}Z\left( y_{d}\pm \varepsilon
\right) $ is the relative jump of $Z$ at the possible point $y_{d}$ of its
periodic discontinuity. Assume first that $Z(y)$ is strictly continuous for
any $y$ (not restricted to $\left[ 0,1\right] $) and hence $\left[ Z\right]
=1.$ Then Eq.\ (\ref{WKB}) yields $\left\vert \Delta _{\mathrm{WKB}%
}^{(0)}\right\vert \leq 1$ and thus can estimate zeros of $\Delta $ but not
the stopbands $\left\vert \Delta \right\vert >1,$ whose widths (the
frequency gaps between cutoffs, see (\ref{19})$_{2,3}$) may well be nonzero
at finite $\omega .$ Thus if $Z(y)$ is continuous then Eq.\ (\ref{WKB})
merely implies that the stopband widths tend to zero at any fixed $k$ as $%
\omega $ tends to infinity. The latter conclusion is also valid even if $\mu
_{2}/\rho $ has periodic jumps but $\rho \mu _{1}$ is continuous throughout,
so that $\left[ Z\right] \neq 1$ indicates existence of nonzero stopbands at
finite $\omega $ but $\left[ Z\right] \rightarrow \left[ Z_{0}\right] =1$ at
$\omega \rightarrow \infty .$ On the other hand, if $\rho \mu _{1}$ does
have a jump and so $\left[ Z_{0}\right] \neq 1,$ then Eq.\ (\ref{WKB}) shows
that the stopband widths remain nonzero as $\omega \rightarrow \infty $.
Having stated this, we hasten to add that a physically sensible profile
model should be related to the frequency $\omega $ in that a finite $\omega $
implies that a probing wave "sees" appropriately abrupt variations of
material properties as jumps, which are of course smoothed out by the
'infinite zoom' of the limit $\omega \rightarrow \infty $. The above WKB
conclusions on the high-frequency trends of cutoffs agree with a less
general framework of, specifically, small periodic perturbations that
provides expressions for the stopband widths through the Fourier series
coefficients, see \cite{AuldI,Brillouin}.

As an example, consider again Fig.\ \ref{fig1}, which is plotted for a
piecewise continuous profile (\ref{0}) that gives $\left[ Z\right] =12\sqrt{%
\left( 1-4k^{2}/\omega ^{2}\right) /\left( 4-k^{2}/\omega ^{2}\right) }$
(note that a 'single periodic discontinuity $y_{d}$' is located at the edges
of the period $T=1$ by (\ref{0}); however, similarly to Remark \ref{13.3}, $%
\Delta _{\mathrm{WKB}}^{\left( 0\right) }$ does not depend on the choice of
the period $\left[ 0,1\right]$ relative to $y_{d}$). It is easy to check
that the exact curves $\Delta $ shown in Fig.\ \ref{fig1}a are well fitted by
the WKB approximation (\ref{WKB}) (not displayed to avoid overloading the
plot) once $\omega $ is greater enough than $k\max \sqrt{\mu _{2}/\rho }=2k$%
. It is also seen from Fig.\ \ref{fig1}a that increasing $\omega $ makes the
curves $\Delta $ for different fixed $k$ tend to that related to $k=0,$ as
predicted by Eq.\ (\ref{WKB}).

In the case of two or more discontinuity points per period, applying the WKB
asymptotics separately along each range of continuity modifies (\ref{WKB})
to the form with two or more phase terms corresponding to the
reflection-transmission at each discontinuity. For more examples of using
the WKB approach to the periodic profile, see \cite{SPG}.

\subsection{Zero-width stopband}

\subsubsection{Complementary definitions of ZWS}

The following definition of a zero-width stopband (ZWS)\footnote{%
It is understood that a ZWS is actually not a 'stopband' (in the sense of
Definition \ref{14.1}). Note that a similar notion of 'zero-width passband'
is inconceivable due to Proposition \ref{19.20}.} is motivated by the
possible occurrence of the second and third cases in (\ref{19.3}).

\begin{definition}
\label{19.5}If $\omega =\omega _{2n}(0,k)=\omega _{2n+1}(0,k)$ or $\omega
=\omega _{2n-1}(\pi ,k)=\omega _{2n}(\pi ,k)$ for some $\omega ,~k\in
\mathbb{R}
$ and $n\in
\mathbb{N}
,$ then this cutoff point $\left( \omega ,k\right) $ is called a ZWS.
\end{definition}

\noindent It is essential that the cutoff curves are analytic (as any $%
\omega _{n}(K,k)$ with fixed $K\in
\mathbb{R}
$ is, see \S 4.1), hence if two of them meet at a point they cannot
conjoin. Thus an isolated ZWS implies intersection of two cutoff curves
on the $(\omega ,k)$-plane and hence a saddle point $\left\vert \Delta
\right\vert =1$ on the Lyapunov-function surface $\Delta (\omega
^{2},k^{2}). $ For the same reason, if, exceptionally (see \S 3.3.3), a ZWS
forms a line $\omega \left( k\right) $ of local extremum $\left\vert \Delta
\right\vert =1$ of $\Delta (\omega ^{2},k^{2}),$ then such line cannot have
an edge point.

A comprehensive account of the properties of ZWS is based on the next
proposition.

\begin{proposition}
\label{19.41}The following statements are equivalent: (i) $(\omega ,k)$ is a
ZWS; (ii) $\Delta (\omega ^{2},k^{2})=\pm 1$ and $\partial \Delta (\omega
^{2},k^{2})/\partial (\omega ^{2})=0$; (iii) $\Delta (\omega ^{2},k^{2})=\pm
1$ and $\partial \Delta (\omega ^{2},k^{2})/\partial (k^{2})=0$; (iv) $%
\mathbf{M}(1,0)=\pm \mathbf{I}$.
\end{proposition}

\noindent \textit{Proof.} The link \textit{(i)}$\Leftrightarrow $\textit{(ii)%
} follows from Definition \ref{19.5} and Proposition \ref{19.20}. The link
\textit{(i)}$\Rightarrow $\textit{(iv)} can be inferred e.g. via (\ref{19.4}%
), which tells us that assuming \textit{(i) }entails $M_{2}(1,0) =M_{3}(1,0)
=0$ and hence $M_{1}(1,0) M_{4}\left( 1,0\right) =\det \mathbf{M}=1,$ where $%
M_{1,}$ $M_{4}$ are real by (\ref{13})$_{2}$. Since \textit{(i)} also means $%
\mathrm{tr}\mathbf{M}\left( 1,0\right) =\pm 2,$ it follows that $\mathbf{M}%
(1,0) =\pm \mathbf{I}$ as stated. Next let us show \textit{(iv)}$\Rightarrow
$\textit{(ii)}. Assume $\mathbf{M}(1,0) =\pm \mathbf{I}$ for some $%
\widetilde{\omega },\widetilde{k}\in
\mathbb{R}
.$ Note that $\Delta (\widetilde{\omega }^{2},\widetilde{k}^{2})=\pm 1$ by (%
\ref{14.0}). The (double) eigenvalue $q=\mathrm{e}^{iK}=\pm 1$ of $\mathbf{M}%
(1,0) =\pm \mathbf{I}$ has geometrical multiplicity 2, hence $\widetilde{%
\omega }^{2}$ is an eigenvalue of $\mathcal{A}_{K}(\widetilde{k})$ of
multiplicity $2$ by Corollary \ref{14.02}. Now consider some $K^{\prime }\in
\mathbb{R}
$ arbitrary close to $K$ that yields $\cos K^{\prime }=\Delta (\omega ^{2},%
\widetilde{k}^{2})\in (-1,1).$ Since $\widetilde{\omega }^{2}$ is a double
eigenvalue of $\mathcal{A}_{K}(\widetilde{k})$, the self-adjoint operator $%
\mathcal{A}_{K^{\prime }}(\widetilde{k})$ has two distinct simple
eigenvalues $\omega ^{2}\left( K^{\prime },\widetilde{k}\right) $ close to $%
\widetilde{\omega }^{2},$ and, by Propositions \ref{14.01} and \ref{19.00},
these are distinct simple zeros of $\Delta (\omega ^{2},\widetilde{k}%
^{2})-\cos K^{\prime }.$ Therefore $\Delta (\widetilde{\omega }^{2},%
\widetilde{k}^{2})=\pm 1$ is a local extremum of $\Delta (\omega ^{2},%
\widetilde{k}^{2}),$ i.e. $\partial \Delta /\partial (\omega ^{2})=0$ at $%
\widetilde{\omega }^{2},~\widetilde{k}^{2},$ which is equivalent to \textit{%
(ii).} Note that reversing the above reasoning proves \textit{(ii)}$%
\Rightarrow $\textit{(iv)}~ without appeal to (\ref{19.4}), and that
invoking $\mathcal{B}_{K}(\omega )$ in place of $\mathcal{A}_{K}(k)$%
~provides a similar proof of \textit{(iii)}$\Leftrightarrow $\textit{(iv) }%
(see also Proposition \ref{24.42} below).\textit{\ }$\blacksquare $

\noindent Note that the point $\omega =0,~k=0$ which yields $\Delta =1$ is
not a ZWS since it does not satisfy any of the above statements, which is
evident from (\ref{19})-(\ref{19.1}).

Proposition \ref{19.41} implies that the multiplicity of $\omega ^{2}$, $%
k^{2}$ as the roots of equation $\Delta (\omega ^{2},k^{2})-\cos K$ at $K\in
\mathbb{R}
$ is the same as their multiplicity as the eigenvalues of $\mathcal{A}%
_{K}(k),~\mathcal{B}_{K}(\omega )$ (this multiplicity is 2 at a ZWS and 1
elsewhere). This is noteworthy since such a parity does not always hold
inside a 'true' stopband $K\notin
\mathbb{R}
,$ where a double root $\omega ^{2}$ or $k^{2}$ of Eq.\ (\ref{14}) is not a
double eigenvalue of, respectively, $\mathcal{A}_{K}(k)$~or $\mathcal{B}%
_{K}(\omega )$ which are no longer self-adjoint for $K\notin
\mathbb{R}
.$ It is also pointed out that the eigenvalue $q=\mathrm{e}^{iK}$ of $%
\mathbf{M}(1,0)$ has an algebraic multiplicity 2 at any cutoff, while its
geometrical multiplicity is 2 only at cutoffs that are ZWS.

\begin{corollary}
\label{19.42}The matrix $\mathbf{M}(1,0)$ is non-semisimple for any cutoff $%
(\omega ,k)$ unless it is a ZWS.
\end{corollary}

\noindent We note that the non-semisimple nature of the monodromy matrix at
the cutoffs has important ramifications for the interpretation of its matrix
logarithm, which has been proposed as the basis for dynamic effective medium
models, see \cite{Shuvalov10c,Shuvalov11}.

\subsubsection{Considerations of the existence of ZWS}

To begin with, it is recalled that the period $T=1$ is everywhere understood
as a \textit{minimal} possible period, so that trivial ZWS which turn up
when $T$ is a multiple of the minimal period are disregarded.

Given an arbitrary periodic $\mathbf{Q}(y),$ the condition $\mathbf{M}%
(1,0)=\pm \mathbf{I}$ stipulating  existence of ZWS imposes three real
constraints on two parameters $\omega ,$ $k$ and hence is unlikely to hold.
However, if the profile $\mathbf{Q}(y)$ is symmetric (even) about the
midpoint of the period $\left[ 0,1\right] $, then, by virtue of (\ref{13a}),
the above condition on $\mathbf{M}(1,0)$ implies only two constraints and
thus such profile can be expected to yield a set of ZWS points
(intersections of cutoff curves $\left\vert \Delta \right\vert =1$) on the $%
\left( \omega ,k\right) $-plane. More precisely, since the cutoffs are
independent of how the period interval is fixed (see Remark \ref{13.3}), ZWS
are expected to exist if a given profile $\mathbf{Q}(y)$ admits such a
choice of the period interval $\left[ y_{0},y_{0}+1\right] \equiv \left[ 0,1%
\right] $ within which $\mathbf{Q}(y)$ is symmetric.

Note that by definition any ZWS is also an intersection of Dirichlet and
Neumann branches (\ref{19.4}) while the inverse is generally not true.
Moreover, in contrast to ZWS, the Dirichlet and Neumann branches and hence
their intersections $\big\{\omega ,k\big\}_{\mathrm{D=N}}$ depend on the
choice of the period interval. For instance, let $\mathbf{Q}(y)$ be
symmetric with respect to a fixed period $\left[ 0,1\right] .$ Then the
Dirichlet and Neumann branches coincide with the cutoff curves and hence any
intersection $\big\{\omega ,k\big\}_{\mathrm{D=N}}$ is a ZWS (see e.g. Fig.\ 1 of\textbf{\ \cite{SPK})}. However, if for a given $\mathbf{Q}(y)=\mathbf{Q}%
(y+1)$ the period is shifted so that $\mathbf{Q}(y)$ is not even about its
midpoint, then a new set $\big\{\omega ,k\big\}_{\mathrm{D=N}}$ includes but
generally does not coincide with the (unchanged) set of ZWS.

As a simple explicit example, consider a periodically bilayered structure
where $\mathbf{Q}(y)$ takes two alternating constant values within two
layers $j=1,2$ that constitute a period $\left[ 0,1\right] $. The monodromy
matrix is given by the standard expression
\begin{equation}
\mathbf{M}(1,0)=\left(
\begin{array}{cc}
\cos \psi _{2}\cos \psi _{1}-\frac{Z_{1}}{Z_{2}}\sin \psi _{2}\sin \psi _{1}
& \frac{\mathrm{i}}{Z_{1}}\cos \psi _{2}\sin \psi _{1}+\frac{\mathrm{i}}{%
Z_{2}}\sin \psi _{2}\cos \psi _{1} \\
\mathrm{i}Z_{2}\cos \psi _{1}\sin \psi _{2}+\mathrm{i}Z_{1}\sin \psi
_{1}\cos \psi _{2} & \cos \psi _{2}\cos \psi _{1}-\frac{Z_{2}}{Z_{1}}\sin
\psi _{2}\sin \psi _{1}%
\end{array}%
\right) ,  \label{M}
\end{equation}%
where $Z_{j}$ is the layer impedance defined in (\ref{3.2}) and $\psi
_{j}=\omega Z_{j}d_{j}/\mu _{1j}$ with $d_{j}$ for the layer thickness. The
set of Dirichlet/Neumann intersections $\big\{\omega ,k\big\}_{\mathrm{D=N}}$
is defined by simultaneous vanishing of both off-diagonal components of (\ref%
{M}), which implies the following three options: (i) $\left\{ \sin \psi
_{1}=0,\ \sin \psi _{2}=0\right\} ,$ (ii) $\left\{ \cos \psi _{1}=0,\ \cos
\psi _{2}=0\right\} $ and (iii) $\left\{ Z_{1}=Z_{2},\ \sin \left( \psi
_{1}+\psi _{2}\right) =0\right\} $, where (iii) may or may not hold for real
$\omega ,k$ \cite{ADL}$.$ It is seen that (i) and (iii) yield $\mathbf{M}%
(1,0)=\pm \mathbf{I}$. Thus (i) and maybe (iii) define ZWS, while (ii) does
not.

Recall that an infinite periodically bilayered structure can always be
considered over a three-layered period where the same stepwise profile $%
\mathbf{Q}(y)$ is symmetric. Hence the fact that any bilayered profile
always admits ZWS (see e.g. Fig.\ 2b in \S 4.1) is consistent with the above
conclusion  that ZWS should  be expected for the profiles $\mathbf{Q}%
(y)$ that can be defined as symmetric over some interval $\left[
y_{0},y_{0}+1\right] .$

\subsubsection{Model examples of regular loci of ZWS}

\begin{itemize}
\item Uniform normal impedance: $Z_{0}^{2}\equiv \rho (y) \mu _{1}(y) =const$
at any $y\in \left[ 0,1\right] .$
\end{itemize}

\noindent Let $k=0.$ The coefficient in (\ref{3.2}) at $k=0$ is $Z\left(
\widetilde{y}\right) =Z_{0}\left( \widetilde{y}\right) ,$ which is constant
at $Z_{0}(y)=const$ by virtue of $\mu _{1}>0.$ Alternatively, note from (\ref%
{10})$_{2}$ that $\mathbf{Q}(y)$ with $k=0$ and $Z_{0}=const$ has constant
eigenvectors. Either of these observations readily shows that, for $k=0$, a
dependence of $\omega $ on $K>0$ (not restricted to $K\in \left[ 0,\pi %
\right] $) is a straight line and thus all stopbands are ZWS, that is, there
is no stopbands at all. The only difference with the case of constant $\rho $
and $\mu _{1}$ is the slope of $\omega \left( K,0\right) $ which is
specified as follows:
\begin{equation}
\omega \left( K,0\right) =KZ_{0}/\left\langle \rho \right\rangle
=K/Z_{0}\left\langle \mu _{1}^{-1}\right\rangle ,  \label{19.6}
\end{equation}

\begin{itemize}
\item Uniform speed: $c^{2}\equiv \mu _{2}(y) /\rho (y) =const$ at any $y\in %
\left[ 0,1\right] $ ($\mu _{1}(y) $ is arbitrary).
\end{itemize}

\noindent The Lyapunov function is then $\Delta (\omega ^{2},k^{2})=\Delta
(\omega ^{2}-c^{2}k^{2},0)$, from (\ref{10})$_{2}$, and consequently
\begin{equation}
\omega _{n}(K,k)=\sqrt{\omega _{n}^{2}\left( K,0\right) +c^{2}k^{2}}.
\label{19.7}
\end{equation}%
Hence if $\omega _{n}^{2}\left( \pi m,0\right) $ with $m=0$ or $1$ is a
zero-width stopband, that is, if $\omega _{n}\left( \pi m,0\right) =\omega
_{n+1}\left( \pi m,0\right) $, then by (\ref{19.7}) $\omega _{n}\left( \pi
m,k\right) =\omega _{n+1}\left( \pi m,k\right) $ $\forall k,$ i.e. the
entire line $\left( \omega _{n}^{2}\left( \pi m,k\right) ,k\right) $ for any
$k\in
\mathbb{R}
$ is a locus of ZWS. Note from (\ref{19.7}) and (\ref{19.1}) that the first
cutoff (which is not a ZWS) is $\omega _{1}\left( 0,k\right) =ck$ $=\omega _{%
\mathrm{N},1}(k)$, where $\omega _{\mathrm{N},1}(k)$ is the first Neumann
solution for $y\in \left[ 0,1\right] $.

\begin{itemize}
\item Uniform normal impedance and speed: $Z_{0}^{2}=const$ and $c^{2}=const$
at any $y\in \left[ 0,1\right] .$
\end{itemize}

\noindent Now Eqs.\ (\ref{19.6}) and (\ref{19.7}) together imply that all
stopbands are ZWS for any $k\in
\mathbb{R}
$. Note that the inverse statement is true under an additional condition of
absolute continuity of $Z_{0},$ by the Borg theorem \cite{Borg}.

\subsection{Explicit expressions for the derivatives of $\Delta $}

\noindent

\begin{theorem}
The derivatives of $\Delta (\omega ^{2},k^{2})$ at any $\omega ^{2},k^{2}\in
\mathbb{C}
\ $(hence in both the passbands and the stopbands at $\omega ^{2},k^{2}\in
\mathbb{R}
$) are given by the formula%
\begin{equation}
\begin{aligned} & \frac{\partial ^{n+m}\Delta (\omega ^{2},k^{2})}{\partial
( \omega ^{2}) ^{n}\partial \left( k^{2}\right) ^{m}} =\frac{1}{2}{\left(
-i\right) ^{n}i^{m}n!m!}\int_{0}^{1}\mathrm{d}\varsigma
_{1}\int_{0}^{\varsigma _{1}}\mathrm{d}\varsigma _{2}\ldots
\int_{0}^{\varsigma _{n+m-1}}\mathrm{d}\varsigma _{n+m} \\ & \qquad \times
F\left( \varsigma _{1},\ldots ,\varsigma _{n+m}\right) M_{2}\left( \varsigma
_{n+m}+1,\varsigma _{1}\right) M_{2}\left( \varsigma _{1},\varsigma
_{2}\right) \ldots M_{2}\left( \varsigma _{n+m-1},\varsigma _{n+m}\right)
,\end{aligned}  \label{20.1}
\end{equation}%
where $M_{2}\left( y_{i},y_{j}\right) $ is a right off-diagonal component of
the matricant $\mathbf{M}\left( y_{i},y_{j}\right) ,$ and%
\begin{equation}
\begin{aligned} \digamma \left( \varsigma _{1},\ldots ,\varsigma
_{n+m}\right) &\equiv \sum_{\sigma \in \Omega }f_{\sigma _{1}}\left(
\varsigma _{1}\right) \ldots f_{\sigma _{n+m}}\left( \varsigma _{n+m}\right)
,\ f_{0}\left( \varsigma \right) \equiv \rho \left( \varsigma \right) ,\
f_{1}\left( \varsigma \right) \equiv \mu _{2}\left( \varsigma \right) ; \\
\Omega &\equiv \left\{ \left( \sigma _{1},\ldots ,\sigma _{n+m}\right) :\
\sigma _{i}=0,1;\ \sum \sigma _{i}=m\ \right\} ,\end{aligned}  \label{20.11}
\end{equation}%
i.e. $\Omega $ is a set of $C_{n+m}^{n}=\left( n+m\right) !/n!m!$
permutations of a set $\left( \sigma _{1},\ldots ,\sigma _{n+m}\right) ,$ in
which each $\sigma _{i}$ is either $0$ or $1$ and their sum is $m$.
\end{theorem}

\noindent \textit{Proof. }The expression (\ref{20.1}) follows from the
following property of matricants of related systems \cite{Pease}: let $\mathbf{Q}%
(y) \mathbf{M}\left( y,y_{0}\right) =\frac{\mathrm{d}}{\mathrm{d}y}\mathbf{M}%
\left( y,y_{0}\right) $ and $\widetilde{\mathbf{Q}}(y) \widetilde{\mathbf{M}}%
\left( y,y_{0}\right) =\frac{\mathrm{d}}{\mathrm{d}y}\widetilde{\mathbf{M}}%
\left( y,y_{0}\right) $ where $\widetilde{\mathbf{Q}}(y) =\mathbf{Q}(y) +%
\mathbf{Q}_{1}(y) ;$ then
\begin{align}
\widetilde{\mathbf{M}}\left( y,y_{0}\right) & =\mathbf{M}\left(
y,y_{0}\right) \widehat{\int }_{y_{0}}^{y}\left[ \mathbf{I}+\mathbf{M}\left(
y_{0},\varsigma \right) \mathbf{Q}_{1}\left( \varsigma \right) \mathbf{M}%
\left( \varsigma ,y_{0}\right) \mathrm{d}\varsigma \right]  \notag
\label{20.2} \\
& =\mathbf{M}\left( y,y_{0}\right) +\int_{y_{0}}^{y}\mathbf{M}\left(
y,\varsigma _{1}\right) \mathbf{Q}_{1}\left( \varsigma _{1}\right) \mathbf{M}%
\left( \varsigma _{1},y_{0}\right) \mathrm{d}\varsigma _{1}+\ldots \\
& +\int_{y_{0}}^{y}\mathrm{d}\varsigma _{1}\ldots \int_{y_{0}}^{\varsigma
_{j-1}}\mathrm{d}\varsigma _{j}\mathbf{M}\left( y,\varsigma _{1}\right)
\mathbf{Q}_{1}\left( \varsigma _{1}\right) \mathbf{M}\left( \varsigma
_{1},\varsigma _{2}\right) \mathbf{Q}_{1}\left( \varsigma _{2}\right) \ldots
\mathbf{M}\left( \varsigma _{j},y_{0}\right) +\ldots .  \notag
\end{align}%
Next note that $\mathbf{Q}\left( y;\omega ^{2},k^{2}\right) \equiv \mathbf{Q}%
\left[ \omega ^{2},k^{2}\right] $ defined by (\ref{10})$_{2}$ is linear in
both $\omega ^{2}$ and $k^{2}.$ Denote small perturbations of $\omega ^{2}$
and $k^{2}$ by $\varepsilon _{\omega }$ and $\varepsilon _{k}$. From (\ref%
{10})$_{2}$,
\begin{equation}
\mathbf{Q}\left[ \omega ^{2}+\varepsilon _{\omega },k^{2}+\varepsilon _{k}%
\right] =\mathbf{Q}\left[ \omega ^{2},k^{2}\right] +i\left( \mu
_{2}\varepsilon _{k}-\rho \varepsilon _{\omega }\right) \mathbf{\Gamma
,\quad \Gamma }=
\begin{pmatrix}
0 & 0 \\
1 & 0%
\end{pmatrix}
.  \label{21}
\end{equation}%
Equation (\ref{20.2}) with $\mathbf{Q}_{1}\equiv i\left( \mu _{2}\varepsilon
_{k}-\rho \varepsilon _{\omega }\right) \mathbf{\Gamma }$ is therefore a
Taylor series of $\widetilde{\mathbf{M}}\equiv \mathbf{M}\left[ \omega
^{2}+\varepsilon _{\omega },k^{2}+\varepsilon _{k}\right] $ about the point $%
\varepsilon _{\omega }=0,$~$\varepsilon _{k}=0$, and hence the derivatives
of the monodromy matrix $\mathbf{M}(1,0) $ with respect to $\omega ^{2}$ and
$k^{2}$ are%
\begin{equation}
\begin{aligned} \frac{\partial ^{n+m}\mathbf{M}(1,0) }{\partial ( \omega
^{2}) ^{n}\partial \left( k^{2}\right) ^{m}}=&\left( -i\right)
^{n}i^{m}n!m!\int_{0}^{1}\mathrm{d}\varsigma _{1}\ldots \int_{0}^{\varsigma
_{n+m-1}}\mathrm{d}\varsigma _{n+m} \\ & \times F\left( \varsigma
_{1},\ldots ,\varsigma _{n+m}\right) \mathbf{M}\left( 1,\varsigma
_{1}\right) \mathbf{\Gamma M}\left( \varsigma _{1},\varsigma _{2}\right)
\mathbf{\Gamma }\ldots \mathbf{M}\left( \varsigma
_{n+m},0\right)\end{aligned}  \label{21.1}
\end{equation}%
with $F$ defined in (\ref{20.11}). Note that $F=\rho \left( \varsigma
_{1}\right) \ldots \rho \left( \varsigma _{n}\right) $ at $m=0$ and $F=\mu
_{2}\left( \varsigma _{1}\right) \ldots \mu _{2}\left( \varsigma _{m}\right)
$ at $n=0.$ Equation (\ref{21.1}) and the definition $\Delta (\omega
^{2},k^{2})=\frac{1}{2}\mathrm{tr}\mathbf{M}(1,0) $ together imply
\begin{equation}
\begin{aligned} & \frac{\partial ^{n+m}\Delta (\omega ^{2},k^{2})}{\partial
( \omega ^{2} ) ^{n}\partial ( k^{2} ) ^{m}} =\frac{1}{2}\frac{\partial
^{n+m} \mathrm{tr}\mathbf{M}( 1,0)} {\partial ( \omega ^{2} ) ^{n}\partial (
k^{2} ) ^{m}} =\frac{ ( -i )
^{n}i^{m}n!m!}{2}\int_{0}^{1}\mathrm{d}\varsigma _{1}\ldots
\int_{0}^{\varsigma _{n+m-1}}\mathrm{d}\varsigma _{n+m} \\ & \qquad \times F
( \varsigma _{1},\ldots ,\varsigma _{n+m} ) \, \mathrm{tr} \big[ \mathbf{M}
( \varsigma _{n+m}+1,\varsigma _{1} ) \mathbf{\Gamma M} ( \varsigma
_{1},\varsigma _{2} ) \mathbf{\Gamma }\ldots \mathbf{M} ( \varsigma
_{n+m-1},\varsigma _{n+m} ) \mathbf{\Gamma } \big] ,\end{aligned}
\label{21.2}
\end{equation}%
where we have used the identity $\mathrm{tr}\left[ \mathbf{M}\left(
1,\varsigma _{1}\right) \ldots \mathbf{M}\left( \varsigma _{n+m},0\right) %
\right] =\mathrm{tr}\left[ \mathbf{M}\left( \varsigma _{n+m},0\right)
\mathbf{M}\left( 1,\varsigma _{1}\right) \ldots \right] $ and the fact that $%
\mathbf{M}\left( \varsigma _{n+m},0\right) =\mathbf{M}\left( \varsigma
_{n+m}+1,1\right) $ due to periodicity. By definition of $\mathbf{\Gamma ,}$
\begin{equation}
\mathbf{M\Gamma }=
\begin{pmatrix}
M_{2} & 0 \\
M_{4} & 0%
\end{pmatrix}
\quad \Rightarrow \quad \mathrm{tr}\left[ \mathbf{M}^{\left( i\right) }%
\mathbf{\Gamma }\ldots \mathbf{M}^{(k) }\mathbf{\Gamma }\right]
=M_{2}^{\left( i\right) }\ldots M_{2}^{(k) },  \label{21.3}
\end{equation}%
which reduces (\ref{21.2}) to the desired form (\ref{20.1}). $\blacksquare $

\begin{corollary}
The first-order derivatives of $\Delta (\omega ^{2},k^{2})$ follow from (\ref%
{20.1}) as
\begin{equation}
\frac{\partial \Delta }{\partial ( \omega ^{2} ) }=\frac{1}{2}%
\int_{0}^{1}\rho (y) m_{2}(y) \mathrm{d}y,\ \frac{\partial \Delta }{\partial
\left( k^{2}\right) }=-\frac{1}{2}\int_{0}^{1}\mu _{2}(y) m_{2}(y) \mathrm{d}%
y,  \label{22}
\end{equation}%
where $im_{2}(y) =$ $M_{2}\left( y+1,y\right) $, see (\ref{13.1}).
\end{corollary}

Interestingly, the expression (\ref{20.1}) for any derivative of $\Delta
(\omega ^{2},k^{2})$ involves, apart from $\rho (y) $ and/or $\mu _{2}(y) $,
only a single, right off-diagonal, element $M_{2}\left( \varsigma
_{i},\varsigma _{j}\right) $ of the matricant. Recall that $\mathop{\rm Re}%
M_{2}=0$ by (\ref{13})$_{2}$, which conforms that (\ref{20.1}) is real as it
must be. Next we will obtain a different representation for the first
derivatives of $\Delta (\omega ^{2},k^{2})$ that is expressed via an
eigenfunction $u(y) $ of (\ref{4}). In contrast to (\ref{20.1}), this
representation is restricted to the passbands $\left\vert \Delta (\omega
^{2},k^{2})\right\vert \leq 1$ and hence to $\omega ^{2},k^{2}\in
\mathbb{R}
$. We note that the components of eigenvectors of $\mathbf{M}\left(
1,0\right) $, which appear in the explicit formulas below, are understood to
be referred to a basis observing the identity (\ref{13}) (an obvious
counterexample is the Jordan form of $\mathbf{M}(1,0) $).

\begin{theorem}
The first derivatives of $\Delta (\omega ^{2},k^{2})$ within the open
passband intervals $\Delta \in (-1,1)$ (and hence $\omega ^{2},k^{2}\in
\mathbb{R}$) satisfy the formulas
\begin{equation}
\frac{\partial \Delta }{\partial (\omega ^{2})}=\frac{\sin K}{\mathbf{w}^{+}%
\mathbf{Tw}}\int_{0}^{1}\rho (y)\left\vert u(y)\right\vert ^{2}\mathrm{d}y,\
\frac{\partial \Delta }{\partial \left( k^{2}\right) }=-\frac{\sin K}{%
\mathbf{w}^{+}\mathbf{Tw}}\int_{0}^{1}\mu _{2}(y)\left\vert u(y)\right\vert
^{2}\mathrm{d}y,  \label{24}
\end{equation}%
where $\mathbf{w}$ is an eigenvector of $\mathbf{M}(1,0)$ corresponding to
the eigenvalue $q=\mathrm{e}^{iK}$, and $u(y)$ is the first component of the
vector $\mathbf{\eta }(y)=\mathbf{M}(y,0)\mathbf{w}=\left( u,i\mu
_{1}u^{\prime }\right) ^{\mathrm{T}}$. At the cutoffs $\Delta =\pm 1,$ Eq.\ (%
\ref{24}) yields zero derivatives in the exceptional case of a ZWS, and is
otherwise modified to
\begin{equation}
\frac{\partial \Delta }{\partial (\omega ^{2})}=\frac{1}{2i\mathbf{w}_{d}^{+}%
\mathbf{Tw}_{g}}\int_{0}^{1}\rho (y)\left\vert u(y)\right\vert ^{2}\mathrm{d}%
y,\ \frac{\partial \Delta }{\partial \left( k^{2}\right) }=-\frac{1}{2i%
\mathbf{w}_{d}^{+}\mathbf{Tw}_{g}}\int_{0}^{1}\mu _{2}(y)\left\vert
u(y)\right\vert ^{2}\mathrm{d}y,  \label{24.1}
\end{equation}%
where $\mathbf{w}_{d}$ and $\mathbf{w}_{g}$ are the proper and generalized
eigenvectors of $\mathbf{M}(1,0)$ that realize its Jordan form (see (\ref{28}%
)), and $u(y)$ is equal to the first component of the vector $\mathbf{\eta }%
(y)=\mathbf{M}(y,0)\mathbf{w}_{d}.$
\end{theorem}

\noindent \textit{Proof of (\ref{24}). }The monodromy matrix $\mathbf{M}%
(1,0) $ at $\left\vert \Delta \right\vert \neq 1$ has distinct eigenvalues $%
q\neq q^{-1}$ and hence linear independent eigenvectors $\mathbf{w}_{1},$ $%
\mathbf{w}_{2}$. Specify their numbering as
\begin{equation}
\mathbf{M}(1,0) \mathbf{w}_{1}=q\mathbf{w}_{1},~\mathbf{M}\left( 1,0\right)
\mathbf{w}_{2}=q^{-1}\mathbf{w}_{2}\mathbf{\ \ }\ \mathrm{with}\ q=\mathrm{e}%
^{iK}\neq q^{-1}=\mathrm{e}^{-iK}.  \label{25.0}
\end{equation}%
According to (\ref{20.2}) and (\ref{21}),
\begin{equation}
\begin{aligned} \frac{\partial \mathbf{M} ( 1,0 ) }{\partial ( \omega ^{2} )
} &=\int_{0}^{1}\mathbf{M} ( 1,y ) \frac{\partial \mathbf{Q} ( y )
}{\partial ( \omega ^{2} ) }\mathbf{M} ( y,0 ) \mathrm{d}y=-i\mathbf{M} (
1,0 ) \int_{0}^{1}\mathbf{P} ( y ) \mathrm{d}y, \\ \mathrm{where}\
\mathbf{P} ( y ) &\equiv \rho ( y ) \mathbf{M}^{-1} ( y,0 ) \mathbf{\Gamma
M} ( y,0 ) \ \ \big( \Rightarrow \mathrm{tr}\mathbf{P} ( y ) =\rho ( y )
\mathrm{tr}\mathbf{\Gamma }=0 \big) .\end{aligned}  \label{20}
\end{equation}%
Hence, the derivative of $\Delta =\frac{1}{2}\mathrm{tr}\mathbf{M}\left(
1,0\right) $ at $\left\vert \Delta \right\vert \neq 1$ is%
\begin{equation}
\frac{\partial \Delta }{\partial ( \omega ^{2}) } =\frac{1}{2i}\left[
q\int_{0}^{1}P_{11}( y) \mathrm{d}y+\frac{1}{q}\int_{0}^{1}P_{22}( y)
\mathrm{d}y\right] =\sin KT\int_{0}^{1}P_{11} (y) \mathrm{d}y,  \label{26}
\end{equation}%
where $P_{11}$ is the upper diagonal element of $\mathbf{P} (y) $ in the
base of vectors $\mathbf{w}_{1}$ and $\mathbf{w}_{2}$. For the passband case
$\Delta \in ( -1,1) $ being considered, the identity $\mathbf{M}^{-1}=%
\mathbf{TM}^{+}\mathbf{T}$ (see (\ref{13})$_{1}$) implies that
\begin{equation}
\mathbf{w}_{1}^{+}\mathbf{Tw}_{2}=0;\ \mathbf{w}_{1}^{+}\mathbf{Tw}_{1},\
\mathbf{w}_{2}^{+}\mathbf{Tw}_{2}\neq 0\ \ \ \ \left[ ( \mathbf{w}_{1}^{+}%
\mathbf{Tw}_{1}) ( \mathbf{w}_{2}^{+}\mathbf{Tw}_2) <0\right] .  \label{25.1}
\end{equation}%
Using (\ref{25.1}), the equality $\mathbf{w}_{1}^{+}\mathbf{TM}^{-1}=$ $%
\left( \mathbf{Mw}_{1}\right) ^{+}\mathbf{T}$ (following from (\ref{13})$_{1}$) and the definition of $\mathbf{\Gamma }$ given in (\ref{21}), we find
that
\begin{equation}
P_{11} (y) \Big\rvert_{\Delta \in ( -1,1) }=\frac{\mathbf{w}_{1}^{+}\mathbf{%
TP} (y) \mathbf{w}_{1}}{\mathbf{w}_{1}^{+}\mathbf{Tw}_{1}}=\frac{\rho (y)
\mathbf{\eta }_{1}^{+} (y) \mathbf{T\Gamma \eta }_{1} (y) }{\mathbf{w}%
_{1}^{+}\mathbf{Tw}_{1}}=\frac{\rho (y) \left\vert u (y) \right\vert ^{2}}{%
\mathbf{w}_{1}^{+}\mathbf{Tw}_{1}},  \label{27}
\end{equation}%
where $\mathbf{\eta }_{1} (y) =\mathbf{M}(y,0) \mathbf{w}_{1}=\left( u,i\mu
_{1}u^{\prime }\right) ^{\mathrm{T}}.$ Based on the numbering in (\ref{25.0}%
) it follows that $\mathbf{\eta }_{1}\left( 1\right) =\mathrm{e}^{iK}\mathbf{%
w}_{1}$ and so $u$ is an eigenfunction of (\ref{4}) (see Corollary \ref%
{14.02}). Substituting (\ref{27}) into (\ref{26}) and setting $\mathbf{w}%
_{1} $ defined in (\ref{25.0}) as $\mathbf{w}_{1}\equiv \mathbf{w}$ leads to
(\ref{24})$_{1}$. The proof of (\ref{24})$_{2}$ is the same. Note that the
sign alternation (\ref{19.2}) of both derivatives at successive cutoffs is
described in (\ref{24}) by the factor $\left( \mathbf{w}^{+}\mathbf{Tw}%
\right) ^{-1}\sin K$ as follows: using $K\in \left[ 0,\pi \right] $ implies $%
\sin K\geq 0$ and alternating sign of $\mathbf{w}^{+}\mathbf{Tw}$ (due to
switching between right- and leftward modes at successive cutoffs); while
using unrestricted $K>0$ implies $\mathbf{w}^{+}\mathbf{Tw}<0$ (rightward
mode) and alternating sign of $\sin K.$ $\blacksquare $

\noindent \textit{Proof of (\ref{24.1})}\textbf{.} Consider a cutoff $\Delta
=\pm 1$ that is not a ZWS and hence implies a non-semisimple $\mathbf{M}%
(1,0).$ Denote%
\begin{equation}
\mathbf{M}(1,0)\mathbf{w}_{d}=q_{d}\mathbf{w}_{d},\quad \mathbf{M}(1,0)%
\mathbf{w}_{g}=q_{d}\mathbf{w}_{g}+\mathbf{w}_{d}\quad \mathrm{at}\quad
\Delta \equiv q_{d}=\pm 1,  \label{28}
\end{equation}%
which defines (not uniquely) the pair $\mathbf{w}_{d}$ and $\mathbf{w}_{g}$
as a basis in which $\mathbf{M}(1,0)$ at $\Delta =\pm 1$ has upper Jordan
form. Hence
\begin{equation}
\frac{\partial \Delta }{\partial (\omega ^{2})}=\frac{1}{2}\mathrm{tr}\frac{%
\partial \mathbf{M}(1,0)}{\partial \left( \omega ^{2}\right) }=\frac{1}{2i}%
\int_{0}^{1}P_{21}(y)\mathrm{d}y,  \label{29}
\end{equation}%
where $P_{21}$ is the left off-diagonal of $\mathbf{P}(y)$ at $\Delta =\pm 1$
in the vector basis of $\mathbf{w}_{d}$ and $\mathbf{w}_{g}.$ The identity $%
\mathbf{M}^{-1}=\mathbf{TM}^{+}\mathbf{T}$ for a non-semisimple $\mathbf{M}%
(1,0)$ implies that%
\begin{equation}
\mathbf{w}_{d}^{+}\mathbf{Tw}_{d}=0;\ \mathbf{w}_{d}^{+}\mathbf{Tw}_{g}\neq
0\ \ \ \left[ \mathop{\rm Re}\mathbf{w}_{d}^{+}\mathbf{Tw}_{g}=0\ \ \mathrm{%
for}\ \det \mathbf{M}=1\right] .  \label{29.1}
\end{equation}%
By (\ref{29.1}) and the definition (\ref{20})$_{2}$ of $\mathbf{P}(y)$,%
\begin{equation}
P_{21}(y)\Big\rvert_{\Delta =\pm 1}=\frac{\mathbf{w}_{d}^{+}\mathbf{TP}(y)%
\mathbf{w}_{d}}{\mathbf{w}_{d}^{+}\mathbf{Tw}_{g}}=\frac{\rho (y)\mathbf{%
\eta }_{d}^{+}(y)\mathbf{T\Gamma \eta }_{d}(y)}{\mathbf{w}_{d}^{+}\mathbf{Tw}%
_{g}}=\frac{\rho (y)\left\vert u(y)\right\vert ^{2}}{\mathbf{w}_{d}^{+}%
\mathbf{Tw}_{g}},  \label{29.2}
\end{equation}%
where $\mathbf{\eta }_{d}(y)=\mathbf{M}(y,0)\mathbf{w}_{d}=\left( u,i\mu
_{1}u^{\prime }\right) ^{\mathrm{T}}.$ Inserting (\ref{29.2}) in (\ref{29})
provides (\ref{24.1})$_{1}$. The proof of (\ref{24.1})$_{2}$ is the same. $%
\blacksquare $

Note that (\ref{24.1}) can also be obtained directly from (\ref{24}) by
taking its limit as $\left\vert \Delta \right\vert <1$ tends to $\left\vert
\Delta \right\vert =\pm 1.$ To do so, proceed from (\ref{25.0}) with $%
q,~q^{-1}$ tending to $q_{d}.$ It is always possible to choose $\mathbf{w}%
_{1},$ $\mathbf{w}_{2}$ so that they have $\mathbf{w}_{d}$ as a common limit
and then $\left( \mathbf{w}_{1}-\mathbf{w}_{2}\right) /\left(
q-q^{-1}\right) $ tends to $\mathbf{w}_{g},$ where $\mathbf{w}_{d}$ and $%
\mathbf{w}_{g}$ satisfy (\ref{28}). By using this limiting definition of $%
\mathbf{w}_{g}$ and the property $\mathbf{w}_{1}^{+}\mathbf{Tw}_{2}=0$ (see (%
\ref{25.1})$_{1}$), the limit of the pre-integral factor in (\ref{24}) with $%
\mathbf{w\equiv w}_{1}$ corresponding to $q=\mathrm{e}^{iK}$ is found to be
\begin{equation}
\frac{\sin K}{\mathbf{w}_{1}^{+}\mathbf{Tw}_{1}}=\frac{q-q^{-1}}{2i\mathbf{w}%
_{1}^{+}\mathbf{T}\left( \mathbf{w}_{1}-\mathbf{w}_{2}\right) }\underset{%
\Delta \rightarrow \pm 1}{\rightarrow }\frac{1}{2i\mathbf{w}_{d}^{+}\mathbf{%
Tw}_{g}}.  \label{29.3}
\end{equation}%
The factor $\mathbf{w}_{d}^{+}\mathbf{Tw}_{g}$ may also be expressed in
terms of the elements $M_{i}(1,0) \equiv M_{i}$ of the matrix $\mathbf{M}%
(1,0) $ which satisfies (\ref{13}). Using (\ref{28}) yields two alternative
forms of this expression as follows:
\begin{equation}
\mathbf{w}_{d}^{+}\mathbf{Tw}_{g}=\frac{\left\vert \mathbf{w}_{d}\right\vert
^{2}M_{2}^{\ast }}{\left\vert M_{1}-q_{d}\right\vert ^{2}+\left\vert
M_{2}\right\vert ^{2}}=\frac{\left\vert \mathbf{w}_{d}\right\vert
^{2}M_{3}^{\ast }}{\left\vert M_{4}-q_{d}\right\vert ^{2}+\left\vert
M_{3}\right\vert ^{2}}.  \label{29.4}
\end{equation}%
If $M_{1},~M_{4}\neq q_{d}$ then $M_{2},~M_{3}\neq 0,$ and so both formulas
in (\ref{29.4}) are equivalent, which follows from $\mathrm{tr}\mathbf{M}%
(1,0) =2q_{d},$ $\det \left[ \mathbf{M}(1,0) -q_{d}\mathbf{I}\right] =0$ and
(\ref{13})$_{2}$. If $M_{1}=q_{d}$ hence $M_{4}=q_{d}$ (or vice versa), then
either $M_{2}=0$ or $M_{3}=0$, as occurs for instance if $\mathbf{Q} (y) $
is even about the midpoint of the period $\left[ 0,1\right] $, see the end
of \S 3.1. Simultaneous vanishing of both $M_{2},~M_{3}$ is ruled out for a
non-semisimple $\mathbf{M}(1,0) $.

In conclusion, the combination of results (\ref{22}) and (\ref{24}), (\ref%
{24.1}) yields the following interesting observation.

\begin{corollary}
The right-hand sides of (\ref{22}) are equal to those of (\ref{24}) in the
passbands $\Delta \in (-1,1),$ and to those of (\ref{24.1}) at the cutoffs $%
\Delta =\pm 1$ (unless the cutoff is a ZWS).
\end{corollary}

\subsection{Properties of the function $m_{2} (y) $}

An important role of the function $m_{2} (y) $ defined in (\ref{13.1}) is
revealed by the fact that, according to (\ref{22}), the first derivative of $%
\Delta (\omega ^{2},k^{2})$ in $\omega ^{2}$ or $k^{2}$ is an integral of $%
m_{2} (y) $ with a positive weight factor $\rho (y) $ or $\mu _{2} (y) .$
Recall also that zeros of $m_{2} (y) $ are the Dirichlet solutions for the
interval $\left[ y,y+1\right] $, see \S 3.1.

\begin{theorem}
\label{24.10} The continuous function $m_{2}(y)=m_{2}(y+1)$ satisfies the
following properties: (i) if $\Delta (\omega ^{2},k^{2})\in (-1,1)$ then $%
m_{2}(y)$ has no zeros for $y\in \lbrack 0,1]$; (ii) if $\Delta (\omega
^{2},k^{2})=\pm 1$ then $m_{2}(y)\geq 0$ for any $y\in \lbrack 0,1]$ or $%
m_{2}(y)\leq 0$ for any $y\in \lbrack 0,1]$; (iii) if $\Delta (\omega
^{2},k^{2})\notin (-1,1)$ and $\omega ^{2},~k^{2}\in
\mathbb{R}
,$ then $m_{2}(y)$ has only finite number of zeros in $[0,1]$.
\end{theorem}

\noindent \textit{Proof.} Consider \textit{(i).} Suppose that $\Delta \in
\left( -1,1\right) $ and there exists $\widetilde{y}$ such that $m_{2}\left(
\widetilde{y}\right) =0.$ Then $\mathbf{M}\left( \widetilde{y}+1,\widetilde{y%
}\right) $ has eigenvalues $m_{1}\left( \widetilde{y}\right) $ and $%
m_{4}\left( \widetilde{y}\right) $($=m_{1}^{-1}\left( \widetilde{y}\right) $
by $\det \mathbf{M}=1$). Therefore, with reference to Remark \ref{13.3}, $%
\Delta =\frac{1}{2}\left[ m_{1}\left( \widetilde{y}\right) +m_{1}^{-1}\left(
\widetilde{y}\right) \right] ,$ where $m_{1}$ according to (\ref{13.1}) is
real (since $\omega ^{2},~k^{2}\in
\mathbb{R}
$ by Lemma (\ref{19.10})). Hence $\left\vert \Delta \right\vert \geq 1,$
which contradicts the initial assumption$.$ The statement \textit{(ii)}
follows from \textit{(i) }and the analyticity of $\Delta (\omega
^{2},k^{2}). $ Consider \textit{(iii).} First note an identity
\begin{equation}
\mathbf{M}^{\prime } ( y+1,y ) =\mathbf{Q} (y) \mathbf{M} ( y+1,y ) -\mathbf{%
M} ( y+1,y ) \mathbf{Q} (y) \ \Rightarrow \ m_{2}^{\prime } (y) = \frac{
m_{1} (y) -m_{4} (y) }{ \mu _{1}(y ) } ,  \label{22.1}
\end{equation}%
where $^{\prime }\equiv \mathrm{d}/\mathrm{d}y$ (if $y$ is a point
discontinuity of a piecewise continuous $\mathbf{Q} (y) ,$ then $\mathrm{d/d}%
y$ is a right or left derivative). Since $\mu _{1}(y) >0$, it follows that $%
m_{2}^{\prime } (y) =0$ iff $m_{4} (y) =m_{1} (y) .$ Now let us suppose the
inverse of \textit{(iii)}, i.e., that $\Delta \notin (-1,1)$ admits the
existence of an infinite set $\{y_{n}\}_{1}^{\infty }$ for which $%
m_{2}(y_{n})=0$. Without loss of generality we may assume that $%
\lim_{n\rightarrow \infty }y_{n}=y_{0}\in \lbrack 0,1]$. Then $%
m_{2}(y_{0})=0 $ and $m_{2}^{\prime }\left( y_{0}\right) =0$. As shown
above, $m_{2}(y_{0})=0$ yields $m_{4}\left( y_{0}\right) =m_{1}^{-1}\left(
y_{0}\right) $ and so we have $\Delta \notin (-1,1)$ for $\Delta =\frac{1}{2}%
\left[ m_{1}(y_{0})+m_{1}^{-1}(y_{0})\right] \notin (-1,1)$ where $m_{1}$ is
real due to $\omega ^{2},~k^{2}\in
\mathbb{R}
.$ It therefore follows that $m_{4}\left( y_{0}\right) =m_{1}^{-1}\left(
y_{0}\right) \neq m_{1}\left( y_{0}\right) .$ According to (\ref{22.1}),
this contradicts $m_{2}^{\prime }\left( y_{0}\right) =0.$ $\blacksquare $

The above result together with Eq.\ (\ref{22}) provides a simple criterion
for a ZWS, which complements Proposition \ref{19.41}.

\begin{proposition}
\label{24.42} The following statements are equivalent: (i) $(\omega ,k)$ is
a ZWS; (ii) $m_{2}(y)=0$ for any $y.$
\end{proposition}

\noindent \textit{Proof.} Assume \textit{(i). }Then ${\mathbf{M}}(1,0)=\pm {%
\mathbf{I}}$ by Proposition \ref{19.41}. Hence by (\ref{13.2}) ${\mathbf{M}}%
(y+1,y)=\pm {\mathbf{I}}$ and so $m_{2}\equiv 0,$ which is \textit{(ii).}
Now assume \textit{(ii). }It requires that $\Delta =\pm 1$ by\textit{\ }%
Theorem \ref{24.10} and yields $\partial \Delta /\partial (\omega ^{2})=0$
by Eq.\ (\ref{22})$_{1}$. According to Proposition \ref{19.41}, $\Delta
(\omega ^{2},k^{2})=\pm 1,$ $\partial \Delta (\omega ^{2},k^{2})/\partial
(\omega ^{2})=0$ implies that $(\omega ,k)$ is a ZWS, which is \textit{(i).}
$\blacksquare $

Interestingly, the function $m_{3}(y),$ whose zeros are the Neumann
solutions for the interval $\left[ y,y+1\right] $, shares some, but not all,
of the properties of $m_{2}(y).$ For instance, $m_{3}(y)$ displays the same
properties\textit{\ (i)}, \textit{(ii)} stated by Theorem \ref{24.10} for $%
m_{2}(y)$ but it does not have the property \textit{(iii)}. The
dissimilarity stems from the fact that (\ref{22.1})$_{1}$ yields $%
m_{3}^{\prime }(y)=\left( \mu _{2}k^{2}-\rho \omega ^{2}\right) \left(
m_{1}-m_{4}\right) ,$ where, in contrast to (\ref{22.1})$_{1}$, the first
factor is not sign-definite. Also the derivatives of $\Delta (\omega
^{2},k^{2})$ are not expressible via $m_{3}(y)$ as they are via $m_{2}(y)$
in (\ref{22}). As a result, Proposition \ref{24.42} does not hold for $%
m_{3}(y)$ in the sense that while it is true that $m_{3}(y)=0$ for any $y$
if $(\omega ,k)$ is a ZWS, the inverse statement is not. An immediate
counter-example is the point $\omega =0,~k=0,$ where $m_{3}(y)=0$ for any $y$
by (\ref{19.1}) but this point is not a ZWS; moreover, the model case $\mu
_{2}(y)/\rho (y)=const\equiv c^{2}$ mentioned in \S 3.3 ensures $m_{3}\equiv
0$ on the whole cutoff line $\omega _{1}\left( 0,k\right) =ck$ (see (\ref%
{19.7})) which has no ZWS points. Thus, the Dirichlet solution $\omega _{%
\mathrm{D},n}(k)$ for $\left[ y,y+1\right] $ does not depend on $y$ only if $%
(\omega _{\mathrm{D},n},k)$ is a zero-width stopband, but the same is not
generally true for the Neumann solutions.

\section{The dispersion surface $\protect\omega _{n}( K,k) $}

In this Section, we address the multisheet surface $\omega _{n}\left(
K,k\right) =\sqrt{\omega _{n}^{2}(K,k)}\left( \geq 0\right) $ which is
defined by Eq.\ (\ref{14}), and study the curves in its cuts taken at
constant $K$, constant $k$ and constant $\omega $.

\begin{remark}
If Eq.\ (\ref{14}) with either $K$ or $k$ or $\omega $ being fixed defines a
differentiable function, then its derivative of any order can be expressed
in terms of partial derivatives of $\Delta ( \omega ^{2},k^{2} ) $ given in (%
\ref{20.1}).
\end{remark}

\noindent Below we examine in detail the first non-zero derivatives. The
higher-order ones are easy to obtain in a similar fashion by differentiating
(\ref{14}). It is understood hereafter that $\omega ,~k\in
\mathbb{R}
.$ By (\ref{14}), $\omega _{n}( K,k) =\omega _{n}\left( -K,k\right) =\omega
_{n}\left( K,-k\right) $ which permits confining considerations to $%
\mathop{\rm Re}K\geqslant 0,~k\geqslant 0.$

\subsection{The function $\protect\omega _{n}(k) $ for fixed $K$}

\begin{figure}[H]
\centering
\includegraphics[width=14cm]{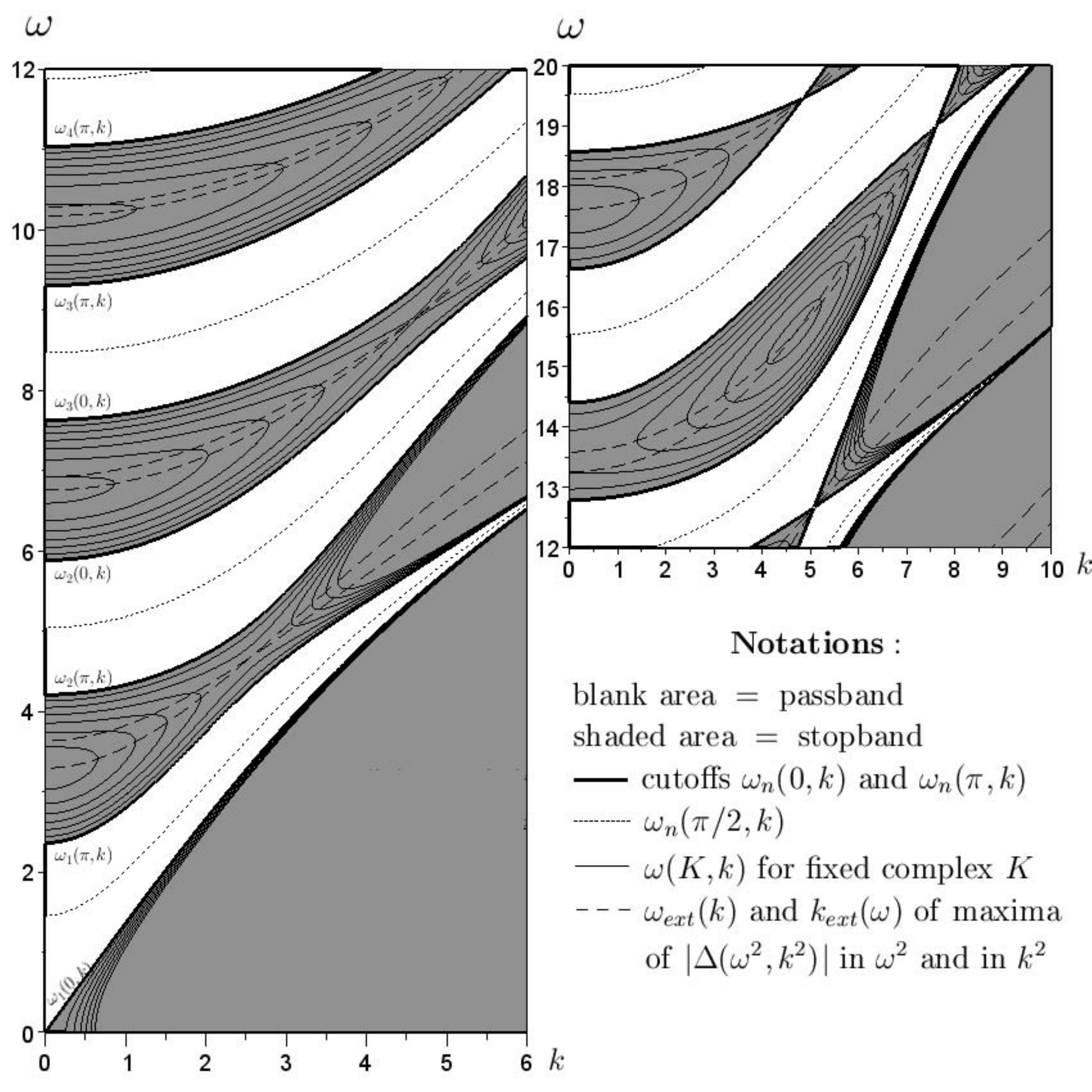}
\caption{(a) (left) The curves $\protect\omega _{n}(k) \equiv \protect\omega
_{n}( K,k) $ at different fixed $K$ for the profile
(\protect\ref{0}). (b) Sections of
the curves for the piecewise constant profile defined by $\protect\mu _{1}=~%
\protect\mu _{2}=1, $ $\protect\rho =1$ for $y\in \lbrack 0,1/2)$ and $%
\protect\mu _{1}=\protect\mu _{2}=12,$ $\protect\rho =2$ for $y\in
(1/2,1]$.} \label{fig2}
\end{figure}

Consider the dependence of $\omega _{n}(k) \equiv \omega _{n}( K,k) $ for
fixed $K$, Fig.\ \ref{fig2}. By Eq.\ (\ref{14}), the branches $\omega _{n}(k) $
are defined as level curves $\Delta (\omega ^{2},k^{2})\left( =\cos K\right)
=const,$ which lie in the passbands for fixed $K\in
\mathbb{R}
\Leftrightarrow $ $\left\vert \Delta \right\vert \leq 1$ and in the
stopbands for fixed complex $K\in \pi
\mathbb{Z}
+i\left(
\mathbb{R}
\backslash 0\right) \Leftrightarrow \left\vert \Delta \right\vert >1$ (note
that the branch numbering (\ref{19.3}) does not apply in the stopbands, see
the discussion of Fig.\ \ref{fig2} below).

\begin{proposition}
If $\omega \neq 0$ and $\partial \Delta /\partial ( \omega ^{2} ) \neq 0,$
then
\begin{equation}
\frac{\mathrm{d}\omega _{n}}{\mathrm{d}k}=\frac{k}{\omega _{n}}\frac{\mathrm{%
d}\omega _{n}^{2}}{\mathrm{d}\left( k^{2}\right) }=-\frac{k}{\omega _{n}}%
\frac{\partial \Delta /\partial \left( k^{2}\right) }{\partial \Delta
/\partial ( \omega ^{2} ) },  \label{30.0}
\end{equation}%
where by (\ref{22}), (\ref{24}) and (\ref{24.1})%
\begin{equation}
\frac{\mathrm{d}\omega _{n}^{2}}{\mathrm{d}\left( k^{2}\right) }=\frac{%
\int_{0}^{1}\mu _{2} (y) m_{2} (y) \mathrm{d}y}{\int_{0}^{1}\rho (y) m_{2}
(y) \mathrm{d}y} \Bigg\rvert_{\substack{ K\in\mathbb{R} \ \mathrm{or}  \\ %
K\in \pi \mathbb{Z} +i \mathbb{R}}} =\frac{\int_{0}^{1}\mu _{2} (y)
\left\vert u_{n} (y) \right\vert ^{2}\mathrm{d}y}{\int_{0}^{1}\rho (y)
\left\vert u_{n} (y) \right\vert ^{2}\mathrm{d}y}\Bigg\rvert_{K\in \mathbb{R}
}.  \label{30}
\end{equation}
\end{proposition}

In addition,
\begin{equation}
\frac{\mathrm{d}\omega _{1}}{\mathrm{d}k} \Big\rvert_{\substack{ \omega=0
\\ k=0}} =\sqrt{\frac{\left\langle \mu _{2}\right\rangle }{\left\langle \rho
\right\rangle }} ;\qquad \frac{\mathrm{d}\omega _{n}}{\mathrm{d}k} %
\Big\rvert _{\substack{ \omega\ne 0  \\ k=0}} =0 ,\qquad \frac{\mathrm{d}k}{%
\mathrm{d}\omega _{1}} \Big\rvert_{\substack{ \omega=0  \\ k\ne 0}} =0.
\label{32}
\end{equation}

\noindent The former equality follows from (\ref{19}) or else from (\ref{30}%
) where $m_{2} (y) $ and $u_{1} (y) $ are constant at\ $\omega ,\ k=0$ in
view of (\ref{19.1}). The two other equalities in (\ref{32}) follow from (%
\ref{30.0}) and $\mathrm{d}\omega _{n}^{2}/\mathrm{d}\left( k^{2}\right)
\neq 0$ (note that $\omega =0,\ k\neq 0$ belongs to the stopband area where (%
\ref{30})$_{1}$ applies, see Fig.\ \ref{fig2}a).

For $K\in
\mathbb{R}
,$ the excluded case $\partial \Delta /\partial (\omega ^{2})=0$  in (\ref%
{30.0}) is related to ZWS discussed in \S 3.3. According to Proposition \ref%
{19.41}, if $\partial \Delta /\partial (\omega ^{2})$ at $K\in
\mathbb{R}
$ becomes zero then so does $\partial \Delta /\partial \left( k^{2}\right) $
and their simultaneous vanishing implies a ZWS. Barring extraordinary cases
mentioned in 3.3.3, ZWS is an intersection point $\left( \omega ,k\right) _{%
\mathrm{zws}}$ of two \textit{analytic} curves $\omega _{n}\left( k\right) $
(as rigorously confirmed in Proposition \ref{32.0} below), so there exist
two derivatives at $\left( \omega ,k\right) _{\mathrm{zws}}$. Their values
can be determined by continuity from either of equations\ (\ref{30}) applied
in the vicinity of $\left( \omega ,k\right) _{\mathrm{zws}}.$ Note that Eq.\
(\ref{30})$_{1}$ is not defined strictly at $\left( \omega ,k\right) _{%
\mathrm{zws}}$\textit{\ }(where $m_{2}(y)=0$\textbf{\ }$\forall y$, see
Proposition \ref{24.42})\textbf{\ }while Eq.\ (\ref{30})$_{2}$ is, provided
that $u_{n}(y)$ implies two different eigenfunctions from a subspace
corresponding to two intersecting curves $\omega _{n}\left( k\right) $ at $%
\left( \omega ,k\right) _{\mathrm{zws}}$.

\begin{proposition}
\label{32.0}The curves $\omega _{n}(k) $ for fixed $K\in
\mathbb{R}
$ are monotonically increasing at $k>0$.
\end{proposition}

\noindent \textit{Proof.} The function $\omega _{n}^{2} ( k^{2} ) $ is
analytic for any $K\in
\mathbb{R}
$ since $\mathcal{A}_{K}(k) $ is a family of analytic operators of Kato's
type A \cite{Kato}. Hence if $\partial \omega _{n}^{2}/\partial ( k^{2} ) =0
$ for some real $k^{2},$ then there exists complex $\widetilde{k}^{2}$ in
the vicinity of $k^{2}$ for which $\omega ^{2}=\omega _{n}^{2}\big(%
\widetilde{k}^{2}\big)$ is real. But this would mean that the operator $%
\mathcal{B}_{K}( \omega ) $ has a complex eigenvalue $k^{2}$ equal to $%
\widetilde{k}^{2},$ which is impossible. Thus $\omega _{n}(k) $ at $K\in
\mathbb{R}
$ is a monotonic function. It increases by virtue of (\ref{30})$_{2}$. To
provide a fully self-consistent proof within the operator approach, note
that (\ref{30})$_{2}$ can also be obtained by applying the perturbation
theory \cite{Krein} to $\mathcal{A}_{K}$ given by (\ref{5}), so that%
\begin{equation}
\frac{\mathrm{d}\omega _{n}^{2}}{\mathrm{d} ( k^{2} ) } =\frac{\mathrm{d}~~}{%
\mathrm{d} ( k^{2} ) }\frac{\left( \mathcal{A}_{K}u_{n},u_{n}\right) _{\rho }%
}{\left\Vert u_{n}\right\Vert _{\rho }^{2}}=\frac{1}{\left\Vert
u_{n}\right\Vert _{\rho }^{2}} \big( \frac{\mathrm{d}\mathcal{A}_{K}}{%
\mathrm{d} ( k^{2} ) }u_{n},u_{n}\big)_{\rho } = \frac{\int_{0}^{1}\mu _{2}
(y) \left\vert u_{n} (y) \right\vert ^{2}\mathrm{d}y}{\int_{0}^{1}\rho (y)
\left\vert u_{n} (y) \right\vert ^{2}\mathrm{d}y}.\ \ \blacksquare
\label{31}
\end{equation}

Consider the example plotted in Fig.\ \ref{fig2}. It demonstrates
monotonicity of the curves $\omega _{n}(k)\equiv \omega _{n}(K,k)$ at fixed $%
K\in
\mathbb{R}
$ by tracing the cutoff curves at $K=0,\pi $ ($\Leftrightarrow \left\vert
\Delta \right\vert =1$) and the curves at $K=\pi /2$ ($\Leftrightarrow
\Delta =0$) within the passbands. Figure \ref{fig2} also shows that, by
contrast, the curves $\omega \left( k\right) \equiv \omega (K,k)$ in the
stopbands, i.e. at fixed complex $K\in \pi
\mathbb{Z}
+i\left(
\mathbb{R}
\backslash 0\right) $ ($\Leftrightarrow $ the level curves $\left\vert
\Delta \right\vert =const>1$), may be not monotonic and can take a looped
shape, either semi-closed or even fully closed. Note that the numbering of
such curves cannot be defined by the rule (\ref{19.3}) restricted to the
passbands. A looped shape is due to a vertical tangent at a point where $%
\partial \Delta /\partial (\omega ^{2})=0$ (see (\ref{30.0}), (\ref{30})$_{1}$). In any stopband except the lowest one, there exists a pair of curves $%
\omega _{\mathrm{ext}}(k)$ and $k_{\mathrm{ext}}(\omega ),$ on which $%
\left\vert \Delta (\omega ^{2},k^{2})\right\vert =\cosh \left(
\mathop{\rm
Im}K\right) $ has maxima in $\omega ^{2}$ and in $k^{2}$ (in $k$ at $k\neq 0$%
), respectively. Hence each stopband except the lowest must contain looped
curves $\omega \left( k\right) $ with a vertical tangent as they cross the
curve $\omega _{\mathrm{ext}}(k)$ - unless the latter fully merges with $k_{%
\mathrm{ext}}(\omega )$ as in the model case $\mu _{2}(y)/\rho (y)=const$
mentioned in \S 3.3. The curves $\omega _{\mathrm{ext}}(k)$ and $k_{\mathrm{%
ext}}\left( \omega \right) $ may intersect within a given stopband thus
indicating a saddle point or an absolute extremum of $\Delta (\omega
^{2},k^{2})$ (the latter is exemplified in Fig.\ \ref{fig2}b, see the family
of closed level curves $\left\vert \Delta \right\vert >1$). At the same
time, $\omega _{\mathrm{ext}}(k)$ and $k_{\mathrm{ext}}(\omega )$ cannot
contact the cutoff curves except at the point of a ZWS (see Fig.\ \ref{fig2}%
b), which is always a saddle point of $\Delta (\omega ^{2},k^{2})$.

It is shown in Appendix A3 that the lower bound for the branches $\omega
_{n}(k) $ at $K\in
\mathbb{R}
$ is $\min_{y\in \left[ 0,1\right] }\sqrt{\mu _{2}/\rho }.$ In the remainder
of this subsection we prove that this bound is also a common limit of $%
\omega _{n}(k) .$ To do so, it is convenient to introduce the velocity $%
v_{n}=\omega _{n}/k.$ First we specify the derivative of $v_{n}(k) $ in
order to demonstrate its monotonicity (note that it is easy to similarly
obtain sign-definite derivatives at fixed $K\in
\mathbb{R}
$ for any other optional choice of the pair of spectral parameters among $%
\omega ,~k~$and $v$ or $s=v^{-1}$).

\begin{lemma}
Let $K\in
\mathbb{R}
$, $n\in \mathbb{%
\mathbb{N}
}$ be fixed. Then $v_{n}^{2} ( k^{2} ) \equiv \omega _{n}^{2}\left(
k^{2}\right) /k^{2}$ is a decreasing function with derivative
\begin{equation}
\frac{\mathrm{d}v_{n}^{2}}{\mathrm{d} ( k^{2} ) }=-\frac{1}{k^{4}}\frac{%
\int_{0}^{1}\mu _{1}|u_{n}^{\prime } (y) |^{2}\mathrm{d}y}{\int_{0}^{1}\rho
|u_{n} (y) |^{2}\mathrm{d}y}<0,  \label{31.1}
\end{equation}%
where $u_{n}$ and $u_{n}^{\prime }$ are defined by $\mathbf{\eta }(y) \equiv
\left( u,~i\mu _{1}u^{\prime }\right) ^{\mathrm{T}}=\mathbf{M}(y,0) \mathbf{w%
}$ taken at $\omega _{n}^{2}$ (cf. (\ref{24})).
\end{lemma}

\noindent \textit{Proof.} Multiply Eq.\ (\ref{1}) by $u$ $(=u_{n})$,
integrate by parts and divide the result by $k^{2}$, to yield
\begin{equation}
v_{n}^{2}\int_{0}^{1}\rho |u_{n}(y)|^{2}\mathrm{d}y=\frac{1}{k^{2}}%
\int_{0}^{1}\mu _{1}|u_{n}^{\prime }|^{2}\mathrm{d}y+\int_{0}^{1}\mu
_{2}|u_{n}|^{2}\mathrm{d}y.  \label{31.3}
\end{equation}%
Substituting from (\ref{31.3}) along with (\ref{31}) into $\mathrm{d}\omega
_{n}^{2}/\mathrm{d}(k^{2})=k^{2}\mathrm{d}v_{n}^{2}/\mathrm{d}%
(k^{2})+v_{n}^{2}$ leads to (\ref{31.1}). The same result follows by
applying the perturbation theory \cite{Krein} similarly as in (\ref{31}), whence
$\mathrm{d}v_{n}^{2}/\mathrm{d}(k^{2})=-((\mu _{1}u_{n}^{\prime })^{\prime
},u_{n})_{\rho }/k^{4}\left\Vert u_{n}\right\Vert _{\rho }^{2}$ and
integrating by parts yields (\ref{31.1}). $\blacksquare $

\begin{proposition}
\label{31.0}Let $K\in
\mathbb{R}
$ be fixed. Then for any $n\in \mathbb{%
\mathbb{N}
}$%
\begin{equation}
\lim_{k\rightarrow \infty }\frac{\omega _{n}^{2}}{k^{2}}=\min_{y\in \left[
0,1\right] }\frac{\mu _{2} (y) }{\rho (y) }.  \label{31.4}
\end{equation}
\end{proposition}

\noindent \textit{Proof. }Rewrite (\ref{1}) in the form
\begin{equation}
-(\mu _{1}u^{\prime })^{\prime }+k^{2}\left( \frac{\mu _{2}}{\rho }-\frac{%
\omega ^{2}}{k^{2}}\right) \rho u=0.  \label{31.5}
\end{equation}%
where $v^{2}=\omega ^{2}/k^{2}$. For any fixed $v\equiv \alpha >\min \sqrt{%
\mu _{2}/\rho },$ the coefficient $\left( \mu _{2}/\rho \right) -v^{2}$
changes sign on the interval $[0,1]$ and hence there exist infinitely many
distinct values $k^{2}>0$ which satisfy (\ref{31.5}) (see more in \cite{Glazman})%
{\small .} The latter means that any curve $v_{n}(k) ,$ $n\in \mathbb{%
\mathbb{N}
},$ intersects the line $\alpha (k)\equiv \alpha $ for any $\alpha >\min
\sqrt{\mu _{2}/\rho }.$ Combining this statement with the above-mentioned
facts that all $v_{n}(k) $ are decreasing and have the lower bound $\min
\sqrt{\mu _{2}/\rho }$ yields (\ref{31.4}). $\blacksquare $

It is noteworthy that there is no common limit for a finite spectrum of
eigenvalues of a discrete Schr\"{o}dinger operator with a large potential
\cite{KoKu}.

\begin{figure}[H] \centering
 \parbox[b]{0.49\textwidth}{\centering
 \includegraphics{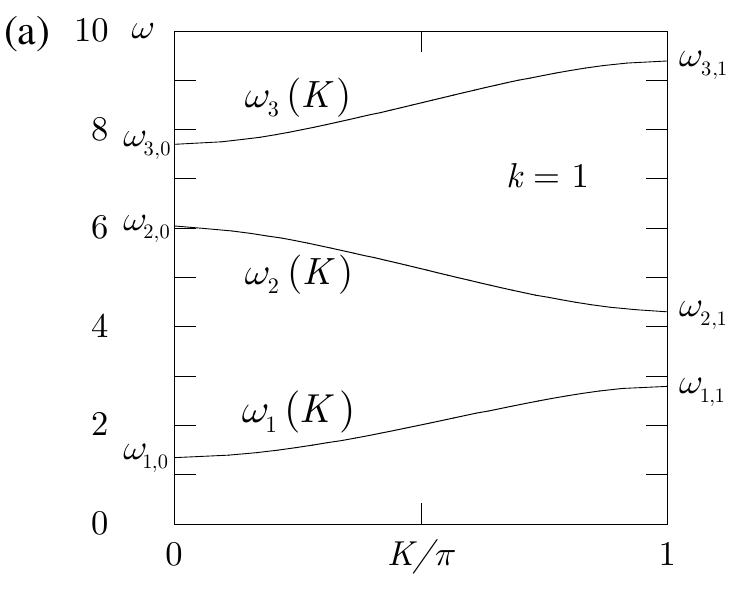}
} \hfil \hfil
\begin{minipage}[b]{0.49\textwidth}
\centering 
\includegraphics{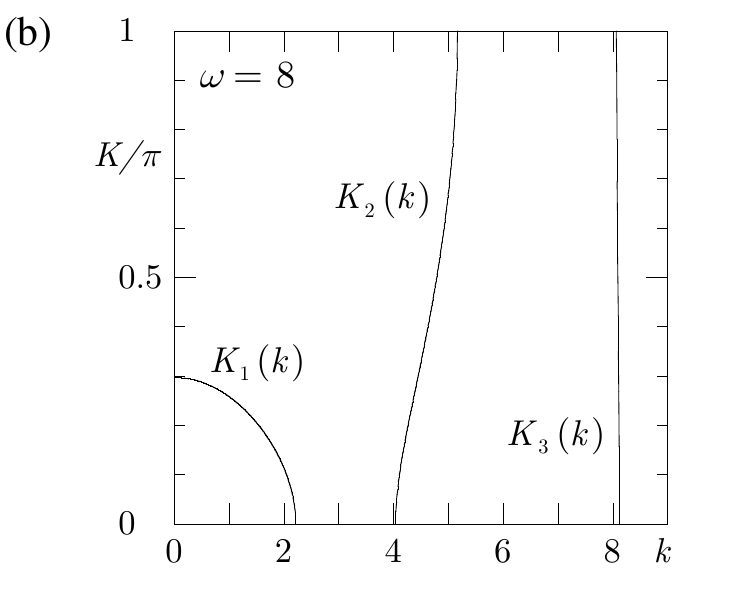}
\end{minipage}
\caption{The Floquet branches $\protect\omega _{n}(K)\equiv \protect\omega %
_{n}(K,k)$ at fixed $k=1.$ (b) Real isofrequency branches $K_{j}(k)$
at fixed $\protect\omega =8$. The same profile (\protect\ref{0}) is
used. The cutoff values of $\protect\omega $ in (a) and of $k$ in
(b) can be compared with Figs.\ \protect\ref{fig1} and
\protect\ref{fig2}a.}\label{fig3}
\end{figure}

\subsection{Function $\protect\omega _{n}(K)$ for fixed $k$}

Consider the function $\omega _{n}(K)\equiv \omega _{n}\left( K,k\right) $
implicitly defined by Eq.\ (\ref{14}): $\Delta (\omega ^{2},k^{2})=\cos K$
at fixed $k$. Since $\omega _{n}(k)$ is periodic and even, it suffices to
deal with one-half of the Brillouin zone $\mathop{\rm Re}K\in \left[ 0,\pi %
\right] $, see Fig.\ \ref{fig3}a. For brevity, denote the cutoff values $%
\omega _{n}\left( \pi m,k\right) $ of $\omega _{n}\left( K,k\right) $ as%
\begin{equation}
\omega _{n}\left( \pi m,k\right) \equiv \omega _{n,m},\ \ \ m=0,1.
\label{33}
\end{equation}%
Let us indicate the passbands and stopbands of $\omega _{n}(K,k)$ by $%
\mathop{\rm Im}K=0$ and $\mathop{\rm Im}K\neq 0$, respectively (the latter
being short for $K=\pi m+i\mathop{\rm Im}K\neq \pi m$). Explicit expressions
for the first non-zero derivative of $\omega _{n}(K)$ readily follow by
expanding both sides of (\ref{14}) and invoking the formulas for $\partial
\Delta /\partial (\omega ^{2})$ obtained in \S 3.4. Note that Eq.\ (\ref{34}%
) with (\ref{24})$_{1}$ for real $K$ (see below) can also be obtained by
means of perturbation theory \cite{Krein} applied to an appropriately modified
form of (\ref{1}), (\ref{2}) with an operator explicitly dependent upon $K$.

\begin{proposition}
If either (i) $\mathop{\rm Im}K=0$ and $K\neq \pi m$ (hence $\partial \Delta
/\omega \neq 0$ by Proposition \ref{19.00}) or (ii) $\mathop{\rm Im}K\neq 0$
and $\partial \Delta /\partial \omega \neq 0,$ then
\begin{equation}
\frac{\mathrm{d}\omega _{n}}{\mathrm{d}K}=-\frac{\sin K}{\left( \partial
\Delta /\partial \omega \right) _{\omega _{n}}},  \label{34}
\end{equation}%
where $\sin K=\sqrt{1-\Delta ^{2}}$ and$\ \partial \Delta /\partial \omega
=2\omega \partial \Delta /\partial ( \omega ^{2} ) $ is given by (\ref{22}$%
_{1}$) or (\ref{24})$_{1}$ for (i) and by (\ref{22})$_{1}$ for (ii). If $%
K=\pi m$ and $\partial \Delta /\partial \omega \neq 0,$ then
\begin{equation}
\frac{\mathrm{d}\omega _{n}}{\mathrm{d}K}=0,\quad \frac{\mathrm{d}^{2}\omega
_{n}}{\mathrm{d}K^{2}}=\frac{\left( -1\right) ^{m+1}}{\left( \partial \Delta
/\partial \omega \right) _{\omega _{n,m}}},\   \label{35}
\end{equation}%
where $\omega _{n,m}=\omega _{n,m}(k) $ are the roots of equation $\Delta (
\omega ^{2},k^{2} ) =\left( -1\right) ^{m}$ and $\partial \Delta /\partial
\omega $ is given by (\ref{22})$_{1}$ or (\ref{24.1})$_{1}$.
\end{proposition}

Consider the special cases where $\partial \Delta /\partial \omega =0$. Let $%
K=\pi m$ and $\partial \Delta /\partial \omega =0$ at $\omega \neq 0,$ which
implies a cutoff $\omega _{n,m}$ corresponding to a ZWS. Then%
\begin{equation}
\frac{\mathrm{d}\omega _{n}}{\mathrm{d}K}={\left( -1\right) ^{m+n+1}}\big/%
\sqrt{\left( -1\right) ^{m+1}\left( \partial ^{2}\Delta /\partial \omega
^{2}\right) _{\omega _{n,m}}}.  \label{37}
\end{equation}%
Next let $\mathop{\rm Im}K\neq 0$ and $\partial \Delta /\partial \omega =0,$
which defines the point $\omega \equiv \omega _{\mathrm{ext}}$ in a stopband
at which $\left\vert \Delta (\omega )\right\vert =\cosh \left(
\mathop{\rm
Im}K\right) $ reaches its maximum $\left\vert \Delta _{\mathrm{ext}%
}\right\vert >1$ (see Fig.\ \ref{fig2} and its discussion in \S 4.1). The
function $\mathop{\rm Im}K(\omega )$ satisfies $\left( \mathrm{d}%
\mathop{\rm
Im}K/\mathrm{d}\omega \right) _{\omega _{\mathrm{ext}}}=0$ and
\begin{equation}
\frac{\mathrm{d}^{2}\mathop{\rm Im}K}{\mathrm{d}\omega ^{2}}=(-1)^{m}\,\frac{%
\left( \partial ^{2}\Delta /\partial \omega ^{2}\right) _{\omega _{\mathrm{%
ext}}}}{\sqrt{\Delta _{\mathrm{ext}}^{2}-1}}\quad \left( <0\ \mathrm{for}%
\mathop{\rm Im}K>0\right) .  \label{38}
\end{equation}%
The explicit form of $\partial ^{2}\Delta /\partial \omega ^{2}$, which
appears in (\ref{37}), (\ref{38}) and is negative at $m=0$ and positive at $%
m=1$, is defined by (\ref{20.1}). It can be written in the following
equivalent forms
\begin{align}
\frac{\partial ^{2}\Delta }{\partial \omega ^{2}}& =4\omega ^{2}\frac{%
\partial ^{2}\Delta }{\partial (\omega ^{2})^{2}}=-4\omega ^{2}\int_{0}^{1}%
\mathrm{d}y\int_{0}^{y}\rho (y)\rho \left( y_{1}\right)
M_{2}(y_{1}+1,y)M_{2}\left( y,y_{1}\right) \mathrm{d}y_{1}  \notag
\label{39} \\
& =-2\omega ^{2}\int_{0}^{1}\mathrm{d}y\int_{y}^{y+1}\rho (y)\rho \left(
y_{1}\right) M_{2}\left( y+1,y_{1}\right) M_{2}(y_{1},y)\mathrm{d}y_{1} \\
& =-2\omega ^{2}\int_{0}^{1}\mathrm{d}y\int_{0}^{1}\rho (y)\rho \left(
y+y_{1}\right) M_{2}\left( y+1,y+y_{1}\right) M_{2}(y+y_{1},y)\mathrm{d}%
y_{1},  \notag
\end{align}%
where $\partial \Delta /\partial \omega =0$ and $\omega \neq 0$ (i.e. $%
\partial \Delta /\partial (\omega ^{2})=0$) have been used. Finally,
consider the case $\omega =0,$ which implies $\partial \Delta /\partial
\omega =0,\ \partial ^{2}\Delta /\partial \omega ^{2}=2\partial \Delta
/\partial (\omega ^{2})$. If both $\omega =0$ and $k=0$ ($\Rightarrow K=0$),
then referring to (\ref{19}), the derivative (\ref{37}) for $m=1$ reduces to
\begin{equation}
\frac{\mathrm{d}\omega _{1}}{\mathrm{d}K}=1\big/\sqrt{\left\langle \rho
\right\rangle \left\langle \mu _{1}^{-1}\right\rangle }.\   \label{36}
\end{equation}%
If $\omega =0$ and $k>0$ ($\Rightarrow K=i\mathop{\rm Im}K\neq 0$), then $%
\left( \mathrm{d}\mathop{\rm Im}K/\mathrm{d}\omega \right) _{\omega =0}=0$
and (\ref{38}) becomes
\begin{equation}
\frac{\mathrm{d}^{2}\mathop{\rm Im}K}{\mathrm{d}\omega ^{2}}=\frac{2\left[
\partial \Delta /\partial (\omega ^{2})\right] _{\omega =0}}{\sqrt{\Delta
^{2}\left( 0,k^{2}\right) -1}},  \label{36.1}
\end{equation}%
where $\left[ \partial \Delta /\partial (\omega ^{2})\right] _{\omega =0}<0$
is given by (\ref{22})$_{1}$.

It is evident from Eq.\ (\ref{34}) that the Floquet branches $\omega _{n}(K)$
for any fixed real $k$ are monotonic in $K\in \left[ 0,\pi \right] $. For
completeness, let us also mention two important results from the general
theory of Schr\"{o}dinger equation \cite{Krein,MO,KaKo} that extend to the case
of Eq.\ (\ref{1}) with fixed $k.$ These results state that $\mathop{\rm Im}%
K(\omega )$ is a convex function and that each branch $\omega _{n}(K)$ has
one and only one inflection point in $K\in \left[ 0,\pi \right] $, unless it
is the lowest branch $\omega _{1}(K)$ at $k=0$ or a branch bounded by a ZWS
at either or both cutoffs $K=\pi m,$ in which case there is no inflection
points. Note in conclusion that Eqs.\ (\ref{35}) and (\ref{37}) provide an
explicit definition for the near-cutoff asymptotics of branches $\omega
_{n}(K)$ that were analyzed in \cite{Craster10a} by a different means (the scaling
approach, also extended in \cite{Craster10a} to 2D-periodicity).

\subsection{The function $K(k) $ for fixed $\protect\omega$}

Consider the dependence of $K(k) =\arccos \Delta (\omega ^{2},k^{2})$ on $%
k\geq 0$ at fixed $\omega .$ Let the branches $K_{j}(k) \in \left[ 0,\pi %
\right] $ for real $K$ be numbered in the order of increasing $k.$ Since $%
\omega _{n}\left( k\right) \equiv \omega _{n}( K,k) $ is strictly increasing
in $k$ (see Fig.\ \ref{fig2}), the number of real branches $K_{j}(k) $ at any
fixed value $\omega $ is fully defined by its position with respect to the
frequency-cutoff points at $k=0$: there is a single real branch $K_{1}(k) $
for a fixed $\omega $ in the interval $0<\omega <\omega _{2}\left( \pi
,0\right) ;$ two real branches $K_{1}(k) ,$~$K_{2}(k) $ for $\omega $ in $%
\omega _{2}\left( \pi ,0\right) <\omega <\omega _{3}\left( 0,0\right) $ ...
etc. Besides, the first real branch $K_{1}\left( k\right) $ starts at $k=0$
and spans a range $[0,\pi )$ or $(0,\pi ]$ iff $\left\vert \Delta ( \omega
^{2},0) \right\vert <1,$ i.e. iff the given $\omega $ is fixed within the
passband at $k=0.$ For example, the value $\omega =8\in \left( \omega _{3}(
0,0) ,\omega _{4}( \pi ,0) \right) $ in Fig.\ \ref{fig2} yields three real
branches $K_{j}(k) $ with $K_{1}(k) \in \lbrack 0,\pi ),$ see Fig.\ \ref{fig3}%
b.

Denote by
\begin{equation}
k_{j,m}( \omega ) \equiv k_{j,m},\ \ \ m=0,1,  \label{44.0}
\end{equation}%
the roots of equation $\Delta ( \omega ^{2},k^{2} ) =\left( -1\right) ^{m}$
which define the points at which $K_{j}(k) =\pi m $ and the given $\omega $
is the cutoff; these points $k_{j,m}$ are separated by the stopband
intervals $\left\vert \Delta \right\vert >1$ where $\mathop{\rm Im}K\neq 0.$
The explicit form of the first derivative of $K\left( k\right) $ for real or
complex $K$ follows from (\ref{14}) and the formulas for $\partial \Delta
/\partial ( k^{2} ) $ in exactly the same way as that $\omega _{n}(k) $ in
\S 4.2.

\begin{proposition}
If $K\neq \pi m$ and $k\neq 0,\ $then%
\begin{equation}
\frac{\mathrm{d}K}{\mathrm{d}k}=-\frac{\partial \Delta /\partial k}{\sin K},
\label{44}
\end{equation}%
where $\partial \Delta /\partial k\neq 0$ for real $K.$ If $K_{j}\left(
k\right) =\pi m$ at $k\neq 0$ and $\left( \partial \Delta /\partial k\right)
_{k_{j,m}}\neq 0,\ $then the locally defined inverse function $k\left(
K\right) $ satisfies%
\begin{equation}
\frac{\mathrm{d}k}{\mathrm{d}K_{j}}=0,\quad \frac{\mathrm{d}^{2}k}{\mathrm{d}%
K_{j}^{2}}=\frac{\left( -1\right) ^{m+1}}{\left( \partial \Delta /\partial
k\right) _{k_{j,m}}}.  \label{45}
\end{equation}%
If $k=0,$ then
\begin{equation}
\frac{\mathrm{d}K_{1}}{\mathrm{d}k}=
\begin{cases}
0,\quad \frac{\mathrm{d}^{2}K_{1}}{\mathrm{d}k^{2}}=-\frac{2}{\sin K_{1}}%
\left[ \partial \Delta /\partial ( k^{2} ) \right] _{k=0} & \mathrm{at}\
K_{1}\neq \pi m, \\
\sqrt{2\left( -1\right) ^{m+1}\left[ \partial \Delta /\partial \left(
k^{2}\right) \right] _{k=0}} & \mathrm{at}\ K_{1}=\pi m.%
\end{cases}
\label{46}
\end{equation}
\end{proposition}

Consider the implication of possibly existing ZWS. Assume that a fixed $%
\omega $ is a ZWS for some $k\neq 0.$ This means that $K_{j}\left(
k_{j,m}\right) =\pi m$ and $\left( \partial \Delta /\partial k\right)
_{k_{j,m}}=0$ where $k_{j,m}\neq 0.$ Then (\ref{45}) is altered to
\begin{equation}
\frac{\mathrm{d}K_{j}}{\mathrm{d}k}=\left( -1\right) ^{m+j}\sqrt{\left(
-1\right) ^{m+1}\left( \partial ^{2}\Delta /\partial k^{2}\right) _{k_{j,m}}}%
.  \label{47}
\end{equation}%
Now assume that a fixed $\omega $ is a ZWS at $k=0$, i.e. let $K_{1}=\pi m$
and $\left[ \partial \Delta /\partial (k^{2})\right] _{k=0}=0.$ Then $%
\mathrm{d}K_{1}/\mathrm{d}k=0$ by (\ref{46}), and%
\begin{equation}
\frac{\mathrm{d}^{2}K_{1}}{\mathrm{d}k^{2}}=\sqrt{2\left( -1\right) ^{m+1}%
\left[ \partial ^{2}\Delta /\partial (k^{2})^{2}\right] _{k=0}}.  \label{48}
\end{equation}%
The second-order derivative of $\Delta $ in (\ref{47}), (\ref{48}) can be
obtained by differentiating (\ref{22})$_{2}$ in the same way as in (\ref{39}%
). Note that $\partial ^{2}\Delta /\partial (k^{2})^{2}$ also appears in the
formula analogous to (\ref{38}) for $\mathrm{d}^{2}\mathop{\rm Im}K/\mathrm{d%
}k^{2}$ at the point $k_{\mathrm{ext}}$ where $\mathrm{d}\mathop{\rm Im}K/%
\mathrm{d}k=0$.

Thus, by (\ref{45}) and (\ref{47}), all real branches $K_{j}(k)$ at fixed $%
\omega $ have vertical tangents at the edge points $K_{j}\left(
k_{j,m}\right) =\pi m,~k_{j,m}\neq 0$ (see Fig.\ \ref{fig1}b), unless the
cutoff $\omega =\omega _{n}\left( \pi m,k_{j,m}\right) $ is a ZWS in which
case $K_{j}(k)$ does not make a right angle with the line $K=\pi m.$ In
turn, by (\ref{46}) and (\ref{48}), the real branch $K_{1}(k)$ has a
horizontal tangent at $k=0,~K\neq \pi $ and a non-zero first derivative at $%
k=0,$ $K=\pi $, unless $\omega =\omega _{n}\left( \pi ,0\right) $ is a ZWS
stopband in which case the slope of $K_{1}(k)$ vanishes at $k=0,$ $K=\pi $.

\begin{remark}
If the cutoff $\omega =\omega _{n}\left( \pi ,0\right) $ is not a ZWS, then
(i) the curve $K_{1}(k)=K_{1}\left( -k\right) $ has a kink at $k=0$; (ii) $%
\mathbf{\nabla }\omega (K,k)=\mathbf{0}$ at $k=0$ by virtue of (\ref{32})
and (\ref{35}).
\end{remark}

\subsection{Convexity of the closed isofrequency branch $K_{1}\left(
k\right) $}

The normal to real isofrequency branches $K_{j}(k)$ defines the direction of
group velocity $\mathbf{\nabla }\omega (K,k)$ which makes their shape
relevant to many physical applications. In particular, negative curvature of
an isofrequency curve is known to give rise to rich physical phenomena
related to wave-energy focussing. Since the function $K(k)=\arccos \Delta $
with $\left\vert \Delta \right\vert \leq 1$ defines a unique $K\in \left[
0,\pi \right] ,$ no vertical line can cross twice the curve $K(k);$ however,
this by itself does certainly not preclude a negative curvature. In fact any
real branch $K_{j}(k),$ which extends from $K_{j}=0$ to $K_{j}=\pi ,$ has
vertical tangents at those edge points and hence must have at least one
inflection between them (unless the exceptional case of ZWS, see \S 4.3).
This simple argument, however, does not apply to the first branch $K_{1}(k)$
if the reference $\omega $ is taken within the passband range at $k=0$ and
hence $K_{1}(k)$ does not reach one of the edge points $0$ or $\pi .$ In
other words, the situation in question is when $K_{1}(k)$ extended by
symmetry to any real $K,~k\lessgtr 0$ forms a \textit{closed} curve.

In the present subsection we address an important case of a relatively low
frequency $\omega $ which is restricted to the passband below the first
cutoff $\omega _{1}\left( \pi ,0\right) $ at the edge of the Brillouin zone $%
K=\pi $ at $k=0.$ For any fixed $\omega <\omega _{1}\left( \pi ,0\right) $,
there is a single real isofrequency branch $K_{1}(k) =\arccos \Delta (
\omega ^{2},k^{2} ) \in \left[ 0,\pi \right) $ that is continuous in the
definition domain $k\in \left[ -k_{1,0},k_{1,0}\right] ,$ where $k_{1,0}$ is
the least root of equation $\Delta =1$ (see (\ref{44.0})). According to (\ref%
{43})$_{1}$,
\begin{equation}
\omega \sqrt{\left\langle \rho \right\rangle /\left\langle \mu
_{2}\right\rangle }\leq k_{1,0}( \omega ) \leq \omega \max\nolimits_{y\in %
\left[ 0,1\right] }\sqrt{\rho (y) /\mu _{2} (y) }.  \label{49}
\end{equation}%
We will show that $K_{1}(k) $ is strictly convex. The proof is preceded by a
lemma.

\begin{lemma}
\label{49.0}For fixed $\omega <\omega _{1}\left( \pi ,0\right) ,$
derivatives of the function $\Delta ( \omega ^{2},k^{2} ) $ of any order in $%
k^{2}$ are strictly positive at $k^{2}\geq 0$.
\end{lemma}

\noindent \textit{Proof.} Let $\omega =0.$ Then $\Delta ( 0,k^{2}) >0$ for $%
k^{2}\geq 0$ by (\ref{A10}) and so $\partial \Delta ( 0,k^{2}) /\partial (
k^{2} ) >0$ for $k^{2}\geq 0$ because $\Delta ( k^{2} ) $ at fixed $\omega
^{2}$ satisfies the conditions of the Laguerre theorem (see Proposition \ref%
{19.20}). In other words, all zeros of $\partial \Delta ( 0,k^{2}) /\partial
( k^{2} ) $ lie in $k^{2}<0$ (see Fig.\ \ref{fig1}b). Now let $0<\omega
<\omega _{1}\left( \pi ,0\right) .$ This means that $-1<\Delta ( \omega
^{2},0) <1$ and so the first zero of $\partial \Delta ( \omega ^{2},k^{2})
/\partial ( k^{2} ) ,$ which is where $\Delta \leq -1,$ still lies in $%
k^{2}<0.$ Thus, if $\omega <\omega _{1}\left( \pi ,0\right) $ then $\partial
\Delta ( \omega ^{2},k^{2}) /\partial ( k^{2} ) >0$ for $k^{2}\geq 0$ and
hence, again by the Laguerre theorem, $\partial ^{p}\Delta /\partial ( k^{2}
) ^{p}>0$ for $k^{2}\geq 0$ and for any $p\geq 1.$ $\blacksquare $

\begin{theorem}
\label{50.0}The curve $K_{1}(k)$ is convex at any fixed $\omega $ such that $%
\omega <\omega _{1}\left( \pi ,0\right) .$
\end{theorem}

\noindent \textit{Proof. }The second derivative of $K_{1}(k) $ is
\begin{equation}
K_{1}^{\prime \prime }(k) =- \left( 1-\Delta ^{2}\right)^{-3/2} h,~\quad
h(k) \equiv \Delta \Big( \frac{\partial \Delta }{\partial k}\Big)^{2}
+\left( 1-\Delta ^{2}\right) \frac{\partial ^{2}\Delta }{\partial k^{2}},
\label{50}
\end{equation}%
where $-1<\Delta ^{2}<1$ for $k\in \left( -k_{1,0},k_{1,0}\right) $, see (%
\ref{44.0})$.$ Note that $\partial \Delta /\partial k=0$ at $k=0$. Let $%
\omega <\omega _{1}\left( \pi ,0\right) $. Then $h(0) =\left( 1-\Delta
^{2}\right) \partial ^{2}\Delta /\partial k^{2}>0$ and $h^{\prime }(k)
=\left( \partial \Delta /\partial k\right) ^{3}+\left( 1-\Delta ^{2}\right)
\partial ^{3}\Delta /\partial k^{3}>0$ according to Lemma \ref{49.0}$.$ Due
to $h(0) >0$ and $h^{\prime }\left( k\right) >0$ at $k>0,$ it follows that $%
h(k) >0$ at $k>0.$ Hence $K_{1}^{\prime \prime }(k) <0$ in its definition
domain $\left[ -k_{1,0},k_{1,0}\right] $. Thus, $K_{1}(k) $ is convex. $%
\blacksquare $

\begin{figure}[H]
\centering
\includegraphics{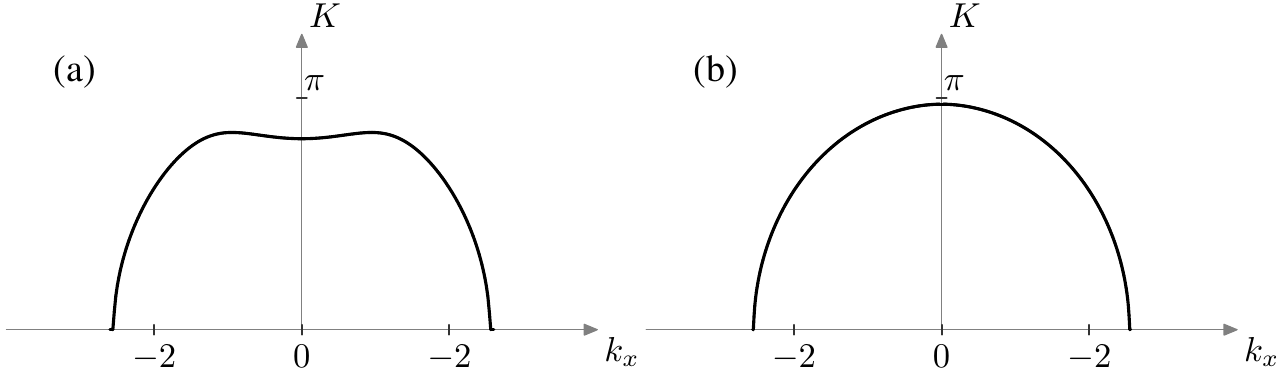}
\caption{(a) The approximate and (b) the exact first isofrequency curve $%
K_{1}(k) =\arccos \left[ \frac{1}{2}\mathrm{tr}\mathbf{M}(1,0)
\right] $ at
fixed $\protect\omega \left( =3.4\right) <\protect\omega _{1}\left( \protect%
\pi ,0\right) $ for a periodically piecewise constant profile defined by $%
\protect\mu _{1}=1,~\protect\mu _{2}=0.35,$ $\protect\rho =0.2$ at
$y\in
\lbrack 0,1/2)$ and $\protect\mu _{1}=0.95,~\protect\mu _{2}=0.4,$ $\protect%
\rho =0.19$ at $y\in (1/2,1].$ The monodromy matrix
(\protect\ref{12}),
which in this case is $\mathbf{M}\left( 1,0\right) =\left( \exp \mathbf{Q}%
_{2}\right) \left( \exp \mathbf{Q}_{1}\right) $ with
$\mathbf{Q}_{j}$ defined by (\protect\ref{10})$_{2}$, is computed
via the series of the co-factor exponentials, keeping four terms for
each of them in the case (a) and 30 terms in the case (b).}
\label{fig4}
\end{figure}

The obtained result sets an important benchmark against any artefacts of
approximate analytical and/or numerical modelling of the first isofrequency
curve $K_{1}(k) =\arccos \Delta ,$ which are possible as a result of
truncating series for $\arccos $ or for $\Delta =\frac{1}{2}\mathrm{tr}%
\mathbf{M}(1,0) $ (see (\ref{12})). Figure \ref{fig4} demonstrates an
example where an approximate computation of $K_{1}(k) $ produces a spurious
concavity. In this regard we note that Figure 1 of \cite{norris92b}, which is
sketch of the generic relation between $K$ and $k$ for fixed but small $%
\omega$, incorrectly gives the suggestion that concavities can occur.

In conclusion, a remark is in order concerning the high-frequency case where
the first isofrequency branch $K_{1}(k) $ defined in $k\in \left[
-k_{1,0},k_{1,0}\right] $ is accompanied by the higher-order branches $%
K_{j\geq 2}(k) .$ In general, $K_{1}(k) $ should stay convex and $K_{j\geq
2}(k) $ should have not more than a single inflection point. However, it
seems possible to construct a theoretical example, though quite peculiar, of
a periodic profile, for which the above is not true.
\medskip

\noindent \textbf{Acknowledgements}

\noindent The authors thank Prof. E. Korotyaev for helpful discussions. AKK
acknowledges support from the University Bordeaux 1 (project AP-2011).



\section*{Appendix}

\subsection*{A1. Properties of the operators $\mathcal{A}_{K}$ and $\mathcal{%
B}_{K}$}

It is evident that the operator $\mathcal{A}_{K}$ defined in (\ref{5}) is
symmetric for $k^{2},~K\in
\mathbb{R}
$, i.e.
\begin{align}
\left( \mathcal{A}_{K}u,v\right) _{\rho }&=-\int_{0}^{1}\left( \mu
_{1}u^{\prime }\right) ^{\prime }v^{\ast }\mathrm{d}y+k^{2}\int_{0}^{1}\mu
_{2}uv^{\ast }\mathrm{d}y=\int_{0}^{1}\mu _{1}u^{\prime }v^{\prime \ast }%
\mathrm{d}y+k^{2}\int_{0}^{1}\mu _{2}uv^{\ast }\mathrm{d}y  \notag \\
&=-\int_{0}^{1}\left( \mu _{1}v^{\prime }\right) ^{\prime \ast }u\mathrm{d}%
y+k^{2}\int_{0}^{1}\mu _{2}uv^{\ast }\mathrm{d}y=\left( u,\mathcal{A}%
_{K}v\right) _{\rho },  \label{A1}
\end{align}%
using the identities $\mu _{1}u^{\prime }v^{\ast }\mid _{0}^{1}=\mu
_{1}v^{\prime }u^{\ast }\mid _{0}^{1}=0$ which follow from the boundary
condition (\ref{6}) on $u,v\in D_{K}$ iff $K$ is real. The proof of the
symmetry of $\mathcal{B}_{K}$ for $\omega ^{2},~K\in
\mathbb{R}
$ is the same.

We now demonstrate that $\mathcal{A}_{K}$ and $\mathcal{B}_{K}$ are
self-adjoint with discrete spectra $\sigma \left( \mathcal{A}_{K}\right)
=\left\{ \omega _{n}^{2}\right\} _{1}^{\infty }$ and $\sigma \left( \mathcal{%
B}_{K}\right) =\left\{ k_{n}^{2}\right\} _{1}^{\infty }$ corresponding to
complete sets of eigenfunctions (as stated in \S 2). This is achieved by
explicit construction of the resolvent of each operator, $\mathcal{R}%
_{K,\lambda }=\left( \mathcal{A}_{K}-\omega ^{2}\right) ^{-1}$ or $\mathcal{R%
}_{K,\lambda }=\left( \mathcal{B}_{K}-k^{2}\right) ^{-1}$, where $\lambda $
implies $\omega ^{2}$ or $k^{2}.$ In order to do so consider the equivalent
equations
\begin{equation}
\begin{array}{c}
\left( \mathcal{A}_{K}-\omega ^{2}\right) u=g,\ \ \omega ^{2}\notin \sigma
\left( \mathcal{A}_{K}\right) \\
\left( \mathcal{B}_{K}-k^{2}\right) u=g,\ \ k^{2}\notin \sigma \left(
\mathcal{B}_{K}\right)%
\end{array}%
\ \ \mathrm{with}\ u (y) \in D_{K},\ g (y) \in L_{\rho ,\mu _{2}}^{2}\left[
0,1\right] ,  \label{A2}
\end{equation}%
which can be recast as
\begin{equation}
\mathbf{\eta }^{\prime } (y) -\mathbf{Q} (y) \mathbf{\eta } (y) =\mathbf{%
\gamma } (y) \ \mathrm{with}\ \mathbf{\gamma } (y) =
\begin{pmatrix}
0 \\
if (y)%
\end{pmatrix}
,\ \mathbf{\eta }\left( 1\right) =\mathrm{e}^{iK}\mathbf{\eta }(0) ,
\label{A3}
\end{equation}%
where $f = -i \rho g$ for $\mathcal{A}_{K}$, $f = i \mu_2g$ for $\mathcal{B}%
_{K}$, and $\mathbf{\eta }$, $\mathbf{Q}$ are defined in (\ref{6}), (\ref{10}%
), respectively. The solution to (\ref{A3}) is a superposition of its
partial solution $\mathbf{\eta }_{p}$ with the solution $\mathbf{\eta }_{0}
(y) $ of the corresponding homogeneous equation:
\begin{equation}
\mathbf{\eta } (y) =\mathbf{\eta }_{p} (y) +\mathbf{\eta }_{0} (y) ,\
\mathbf{\eta }_{p} (y) =\int_{0}^{y}\mathbf{M}\left( y,\varsigma \right)
\mathbf{\gamma }\left( \varsigma \right) \mathrm{d}\varsigma ,\ \mathbf{\eta
}_{0} (y) =\mathbf{M}(y,0) \mathbf{\eta }_{0}(0) \mathbf{.}  \label{A4}
\end{equation}%
The vector $\mathbf{\eta }_{0}(0) $ is found from the quasi-periodic
boundary condition that yields $\mathbf{\eta }_{p}\left( 1\right) +\mathbf{%
\eta }_{0}\left( 1\right) =\mathrm{e}^{iK}\mathbf{\eta }_{0}(0) .$ Thus%
\begin{equation}
\begin{aligned} \mathbf{\eta } (y) &=\int_{0}^{1}\mathbf{G}\left(
y,\varsigma \right) \mathbf{\gamma }\left( \varsigma \right)
\mathrm{d}\varsigma \ \  \mathrm{with} \\ \mathbf{G}\left( y,\varsigma
\right) &=\mathbf{M}\left( y,\varsigma \right) H\left( y-\varsigma \right)
-\mathbf{M}(y,0) \left[ \mathbf{M}(1,0) -\mathrm{e}^{iK}\mathbf{I}\right]
^{-1}\mathbf{M}\left( 1,\varsigma \right) , \end{aligned}  \label{A5}
\end{equation}%
where $H\left( y-\varsigma \right) $ is the Heaviside function and $\mathrm{e%
}^{iK}$ is not an eigenvalue of $\mathbf{M}(1,0) $ for the given $\omega
^{2}\notin \sigma \left( \mathcal{A}_{K}\right) $,\ $k^{2}\notin \sigma
\left( \mathcal{B}_{K}\right) $. It can be checked that the Green-function
tensor $\mathbf{G}\left( y,\varsigma \right) $ satisfies the identity $%
\mathbf{G}\left( y,\varsigma \right) =-\mathbf{TG}^{+}( \varsigma ,y)
\mathbf{T}$, so that its right off-diagonal component satisfies $%
G_{12}\left( y,\varsigma \right) =-G_{12}^{\ast }\left( y,\varsigma \right) $%
. By (\ref{A5})$_{1}$,%
\begin{equation}
u=\mathcal{R}_{K,\lambda }g=\int_{0}^{1}G\left( y,\varsigma ;\lambda \right)
f\left( \varsigma \right) \mathrm{d}\varsigma ,\mathrm{\ where}\ G\left(
y,\varsigma ;\lambda \right) =iG_{12}\left( y,\varsigma \right) .  \label{A6}
\end{equation}%
It is seen that the resolvent $\mathcal{R}_{K,\lambda }$ is an integral
(bounded) self-adjoint operator generated by a piecewise continuous kernel.
The symmetry $\left( \mathcal{R}_{K,\lambda }g,v\right) =\left( g,\mathcal{R}%
_{K,\lambda }v\right) $ follows for any $v\in D_{K}$ from $G\left(
y,\varsigma ;\lambda \right) =G^{\ast }\left( \varsigma ,y;\lambda \right) $
or else from the symmetry of $\mathcal{A}_{K},\ \mathcal{B}_{K}$. Thus $%
\mathcal{R}_{K,\lambda }$ satisfies the Hilbert-Schmidt theorem and $%
\mathcal{A}_{K},$\ $\mathcal{B}_{K}$ therefore possess the above-mentioned
properties.

\subsection*{A2. Bounds of the function $\Delta (\protect\omega ^{2},k^{2})$}

The far-reaching properties of the analytic function $\Delta (\omega
^{2},k^{2})$ stated in Proposition \ref{19.20} follow by applying Laguerre's
theorem to $\Delta (\omega ^{2})$ at any fixed $k^{2}$ and to $\Delta
(k^{2}) $ at any fixed $\omega ^{2}$. A function satisfying Laguerre's
theorem must be an entire function of order of growth less than 2.
Verification of this condition for $\Delta (\omega ^{2},k^{2})$ requires its
uniform estimation in $%
\mathbb{C}
$. The WKB asymptotic expansion (see \S 3.2) is not well-suited for the task
in hand. Here we derive explicit bounds which show that $\Delta (\omega
^{2}) $ and $\Delta (k^{2})$ for, respectively, any $k^{2}$ and $\omega ^{2}$
are entire functions of order of growth
$\frac12$%
. The derivation consists of two Lemmas in which the following auxiliary
notation is used: $f_{\max }\equiv \max f (y) ,~f_{\min }\equiv \min f (y) $
for $f (y) =\rho (y) ,~\mu _{1,2} (y) $ and $y\in \left[ 0,1\right] $.

\begin{lemma}
For any $\omega ,~k\in
\mathbb{C}
,$
\begin{equation}
\left\vert \Delta (\omega ^{2},k^{2})\right\vert \leq \cosh \sqrt{ \mu
_{1\min }^{-1} \big( {\mu _{2\max }\left\vert k\right\vert ^{2}+\rho _{\max
}\left\vert \omega \right\vert ^{2}} \big) }.  \label{A7}
\end{equation}
\end{lemma}

\noindent \textit{Proof}. For any 2$\times $2 matrix $\mathbf{A}$ with the
entries $\left( a_{1}..a_{4}\right) ,$ define $\left\vert \mathbf{A}%
\right\vert $ as
\begin{equation}
\left\vert \mathbf{A}\right\vert =
\begin{pmatrix}
\left\vert a_{1}\right\vert & \left\vert a_{2}\right\vert \\
\left\vert a_{3}\right\vert & \left\vert a_{4}\right\vert%
\end{pmatrix}
\label{A8}
\end{equation}%
and note that $\left\vert \prod\nolimits_{n}\mathbf{A}_{n}\right\vert \leq
\prod\nolimits_{n}\left\vert \mathbf{A}_{n}\right\vert $ where the entrywise
inequality is understood. Recall that $\widehat{\int }$ appearing in (\ref%
{12}) implies a product integral and is an exponential when the integrand
matrix is constant. Hence it follows from (\ref{10})$_{2}$, (\ref{12}) and (%
\ref{14.0}) that
\begin{equation}
\begin{array}{c}
\left\vert \Delta (\omega ^{2},k^{2})\right\vert =\frac{1}{2}\left\vert
\mathrm{tr}\widehat{\int }_{0}^{1}\left[ \mathbf{I}+\mathbf{Q}(y) \mathrm{d}y%
\right] \right\vert =\frac{1}{2}\left\vert \mathrm{tr}\widehat{\int }_{0}^{1}%
\left[ \mathbf{I}+i
\begin{pmatrix}
0 & -\mu _{1}^{-1} (y) \\
\mu _{2} (y) k^{2}-\rho (y) \omega ^{2} & 0%
\end{pmatrix}%
\mathrm{d}y\right] \right\vert \\
\leq \frac{1}{2}\mathrm{tr}\widehat{\int }_{0}^{1}\left[ \mathbf{I}+i
\begin{pmatrix}
0 & \mu _{1\min }^{-1} \\
\mu _{2\max }\left\vert k\right\vert ^{2}+\rho _{\max }\left\vert \omega
\right\vert ^{2} & 0%
\end{pmatrix}
\mathrm{d}y\right] =\cosh \sqrt{\frac{\mu _{2\max }\left\vert k\right\vert
^{2}+\rho _{\max }\left\vert \omega \right\vert ^{2}}{\mu _{1\min }}}\
\blacksquare .%
\end{array}
\label{A9}
\end{equation}

The inequality (\ref{A9}) confirms that $\Delta (\omega ^{2})$ and $\Delta
(k^{2})$ are entire functions with order of growth not greater than
$\frac12$
in each argument. Next we demonstrate that $\Delta $ for certain $\omega
^{2},~k^{2}$ grows no slower than an exponential of power
$\frac12$
of $\omega ^{2}$ and/or $k^{2}$. This will enable us to conclude that the
order of growth of $\Delta (\omega ^{2})$ and $\Delta (k^{2})$ is precisely
$\frac12$%
.

\begin{lemma}
For $\omega ^{2},~k^{2}\in
\mathbb{R}
,$%
\begin{equation}
\left\vert \Delta (\omega ^{2},k^{2})\right\vert \geq \cosh \sqrt{\mu
_{1\max }^{-1} \big( {\mu_{2\min }k^{2}-\rho _{\max }\omega ^{2}}\big) }\
\mathrm{for}\ k^{2}\geq \mu_{2\min }^{-1} {\rho_{\max }} \omega ^{2}.
\label{A10}
\end{equation}
\end{lemma}

\noindent \textit{Proof}. First introduce a class $\mathcal{M}$ of 2$\times $%
2 {matrices such that}
\begin{equation}
\mathcal{M}=\left\{
\begin{pmatrix}
a_{1} & -ia_{2} \\
ia_{3} & a_{4}%
\end{pmatrix}
\right\} ,\ a_{j}\geq 0,\ j=1..4.  \label{A11}
\end{equation}%
For two matrices $\mathbf{A}$ and $\mathbf{B}$ from $\mathcal{M}$, {we} say
that $\mathbf{A\geq }_{\mathcal{M}}~\mathbf{B}$ iff $a_{j}\geq b_{j}$ for
any $j=1..4$. If $\mathbf{A}\in \mathcal{M}$~and $\mathbf{B}\in \mathcal{M}$
then $\mathbf{AB}\in \mathcal{M}$ {also}. Therefore, if $\mathbf{A}_{k},\
\mathbf{B}_{k}\in \mathcal{M}$ and $\mathbf{A}_{k}\geq _{\mathcal{M}}~%
\mathbf{B}_{k}$ for any $k=1..n$ then $\mathbf{A}_{1}..\mathbf{A}_{n}\geq _{%
\mathcal{M}}\mathbf{B}_{1}..\mathbf{B}_{n}$ and $\mathrm{tr}\left( \mathbf{A}%
_{1}..\mathbf{A}_{n}\right) \geq \mathrm{tr}\left( \mathbf{B}_{1}..\mathbf{B}%
_{n}\right) $ ({which is} easy to check for $n=2$ and is therefore valid for
any $n$). We note from (\ref{10})$_{2}$ that $\mu _{2\min }k^{2}\geq \rho
_{\max }\omega ^{2}$ implies $\mathbf{I}+\mathbf{Q} (y) \mathrm{d}y\in
\mathcal{M}$ for any $y\in \left[ 0,1\right] $ and $\mathrm{d}y>0;$ moreover,%
\begin{equation}
\mathbf{I}+\mathbf{Q} (y) \mathrm{d}y\geq _{\mathcal{M}}~\mathbf{I}+i
\begin{pmatrix}
0 & -\mu _{1\max }^{-1} \\
\mu _{2\min }k^{2}-\rho _{\max }\omega ^{2} & 0%
\end{pmatrix}
\mathrm{d}y  \label{A12}
\end{equation}%
and {consequently }
\begin{align}
\Delta (\omega ^{2},k^{2}) &=\frac{1}{2}\mathrm{tr}\widehat{\int }_{0}^{1}%
\left[ \mathbf{I}+\mathbf{Q} (y) \mathrm{d}y\right] \geq \frac{1}{2}\mathrm{%
tr}\widehat{\int }_{0}^{1}\left[ \mathbf{I}+i
\begin{pmatrix}
0 & -\mu _{1\max }^{-1} \\
\mu _{2\min }k^{2}-\rho _{\max }\omega ^{2} & 0%
\end{pmatrix}
\mathrm{d}y\right]  \notag \\
& =\cosh \sqrt{\frac{\mu _{2\min }k^{2}-\rho _{\max }\omega ^{2}}{\mu
_{1\max }}} . \ \ \qquad \blacksquare  \label{A13}
\end{align}
\medskip

\subsection*{A3. Bounds of the first eigenvalue $\protect\omega %
_{1}^{2}(K,k) $}

\begin{proposition}
For $K\in \lbrack -\pi ,\pi ]$ and $k\in
\mathbb{R}
$, the first eigenvalue $\omega _{1}^{2}( K,k) $ is bounded as follows%
\begin{equation}
k^{2}\min_{y\in \left[ 0,1\right] }\frac{\mu _{2} (y) }{\rho (y) }\leq
\omega _{1}^{2}( K,k) \leq \frac{\left\langle \mu _{1}\right\rangle }{%
\left\langle \rho \right\rangle }K^{2}+\frac{\left\langle \mu
_{2}\right\rangle }{\left\langle \rho \right\rangle }k^{2}.  \label{40}
\end{equation}
\end{proposition}

\noindent \textit{Proof.} Let $u_{1}\in D_{K}$ with the unit norm $%
\left\Vert u_{1}\right\Vert _{\rho }=1$ be the eigenfunction vector of $%
\mathcal{A}_{K}$ corresponding to the eigenvalue $\omega _{1}^{2}$. Then%
\begin{equation}
\omega _{1}^{2}=\left( \mathcal{A}_{K}u_{1},u_{1}\right) _{\rho
}=\int_{0}^{1}\mu _{1}\left\vert u_{1}^{\prime }\right\vert ^{2}\mathrm{d}%
y+k^{2}\int_{0}^{1}\mu _{2}\left\vert u_{1}\right\vert ^{2}\mathrm{d}y\geq
k^{2}\int_{0}^{1}\frac{\mu _{2}}{\rho }\rho \left\vert u_{1}\right\vert ^{2}%
\mathrm{d}y\geq k^{2}\min_{y\in \left[ 0,1\right] }\frac{\mu _{2}}{\rho }.
\label{41}
\end{equation}%
An equivalent proof of the lower bound (\ref{41}) follows {by noting} that
the initial equation (\ref{1}) {yields zero as the sum of the positive
operator $-\left( \mu _{1}u^{\prime }\right) ^{\prime }$ and the} operator
multiplying $u$ by $\left( k^{2}\mu _{2}-\omega ^{2}\rho \right) ,$ {%
implying that} the latter factor must be negative. {In order to obtain} the
upper bound, introduce the function $v (y) =\left\langle \rho \right\rangle
\mathrm{e}^{iKy}$ such that $v (y) \in D_{K}$ and $\left\Vert v\right\Vert
_{\rho }=1$. Hence $\omega _{1}^{2}$ as a minimal eigenvalue of $\mathcal{A}%
_{K}$ satisfies%
\begin{equation}
\omega _{1}^{2}=\inf_{u\in D_{K},\ \left\Vert u\right\Vert _{\rho }=1}\left(
\mathcal{A}_{K}u_{1},u_{1}\right) _{\rho }\leq \left( \mathcal{A}%
_{K}v,v\right) _{\rho }=\frac{\left\langle \mu _{1}\right\rangle }{%
\left\langle \rho \right\rangle }K^{2}+\frac{\left\langle \mu
_{2}\right\rangle }{\left\langle \rho \right\rangle }k^{2}.\ \blacksquare
\label{42}
\end{equation}

\begin{corollary}
The bounds of the first cutoff at the centre and the edge of the Brillouin
zone are, respectively,
\begin{equation}
k\min_{y\in \left[ 0,1\right] }\sqrt{\frac{\mu _{2} (y) }{\rho (y) }}\leq
\omega _{1}\left( 0,k\right) \leq k\sqrt{\frac{\left\langle \mu
_{2}\right\rangle }{\left\langle \rho \right\rangle }}; \quad \omega
_{1}\left( 0,k\right) <\omega _{1}\left( \pi ,k\right) \leq \sqrt{\frac{%
\left\langle \mu _{1}\right\rangle }{\left\langle \rho \right\rangle }\pi
^{2}+\frac{\left\langle \mu _{2}\right\rangle }{\left\langle \rho
\right\rangle }k^{2}}.  \label{43}
\end{equation}
\end{corollary}

As stated in Proposition \ref{31.0}, the lower bound (\ref{40}) of $\omega
_{1}( K,k) $ and hence of all curves $\omega _{n}\left( K,k\right) $ {for} $%
K\in
\mathbb{R}
$ is also their limit at $k\rightarrow \infty $. Note that $\omega
_{1}\left( 0,k\right) \geq \omega _{\mathrm{N},1}(k) $ by (\ref{19.4}),
where $\omega _{\mathrm{N},1}(k) $ is the lowest branch of solutions of the
Neumann problem for $y\in \left[ 0,1\right] $. It has the same bounds and
the same limit at $k\rightarrow \infty $ as $\omega _{1}\left( 0,k\right) $.
{In this regard}, recall the model example $\mu _{2} (y) /\rho (y)
=const\equiv c^{2}$ (see \S 3.2), where $\omega _{1}\left( 0,k\right)
=\omega _{\mathrm{N},1}(k) =ck$ merge together with their upper and lower
bounds. By (\ref{43})$_{1}$$,$ unless $\omega _{1}\left( 0,k\right) $ is a
straight line, it has an inflection point (and so does $\omega _{\mathrm{N}%
,1}(k) $). {Furthermore}, the case of constant $\rho ,~\mu _{1,2}$ is an
elementary example of {the equality of} the upper bound in (\ref{40}) and (%
\ref{43})$_{2}$.

\end{document}